\documentclass[preprint]{aastex}
\usepackage{natbib}
\usepackage{graphicx}
\newcommand{\lya}{Ly$\alpha$}
\newcommand{\hi}{H\,{\sc i}}
\newcommand{\hii}{H\,{\sc ii}}
\newcommand{\hst}{\textit{HST}}
\shorttitle{\lya~from $z\sim0.03$ Star-Forming Galaxies}
\shortauthors{Wofford et al.}
\begin{document}
\title{\lya~Escape from $z\sim0.03$ Star-Forming Galaxies: the Dominant Role of Outflows}
\author{Aida Wofford\altaffilmark{1}, Claus Leitherer\altaffilmark{1} and John Salzer\altaffilmark{2}}
\affil{\altaffilmark{1}Space Telescope Science Institute, 3700 San Martin Drive, Baltimore, MD 21218; \email{wofford@stsci.edu}}
\affil{\altaffilmark{2}Astronomy Department, Indiana University, Swain West 408, 727 East Third Street,\\ Bloomington, IN 47405}


\begin{abstract}

The usefulness of \hi~\lya~photons for characterizing star formation in the distant universe is limited by our understanding of the astrophysical processes that regulate their escape from galaxies. These processes can only be observed in detail out to a few $\times100$\,Mpc. Past nearby ($z<0.3$) spectroscopic studies are based on small samples and/or kinematically unresolved data. Taking advantage of the high sensitivity of \hst's COS, we observed the \lya~lines of 20 H$\alpha$-selected galaxies located at $\left<z\right>=0.03$. The galaxies cover a broad range of luminosity, oxygen abundance, and reddening. In this paper, we characterize the observed \lya~lines and establish correlations with fundamental galaxy properties. We find seven emitters. These host young ($\le10\,$Myr) stellar populations, have rest-frame equivalent widths in the range $1-12$\,\AA, and have \lya~escape fractions within the COS aperture in the range $1-12$\,\%. One emitter has a double-peaked \lya~with peaks $370$\,km s$^{-1}$ apart and a stronger blue peak. Excluding this object, the emitters have \lya~and O\,{\sc i} $\lambda$1302 offsets from H$\alpha$ in agreement with expanding shell models and LBG observations. The absorbers have offsets that are almost consistent with a static medium. We find no one-to-one correspondence between \lya~emission and age, metallicity, or reddening. Thus, we confirm that \lya~is enhanced by outflows and is regulated by the dust and \hi~column density surrounding the hot stars.

\end{abstract}


\section{INTRODUCTION}

The potential of \hi~\lya~to identify young galaxies at high redshift was noted over 40 years ago by \cite{par67}.  \lya~corresponds to the $n=1-2$ transition and is in principle the strongest emission line originating in an H\,{\sc ii}~region. In the absence of dust, the \lya/H$\alpha$ line intensity ratio is expected to be in the range $7-12$ for recombination theory cases B-A \citep{ost06}, and the \lya~equivalent width, $EW(\rm{Ly}\alpha)$, is expected to be in the range $100-300$\,\AA~for metallicities between $Z=0.020-0.004$, a normal initial mass function, a constant star formation rate of 1 M$_\odot/$yr$^{-1}$, and ages below 10 Myr \citep{ver08}. In the case of a single burst of star formation, the range of $EW(\rm{Ly}\alpha)$ is $\sim 0-300$ under the same conditions. A few times $10^7-10^9$ yr after a burst of star formation, $EW(\rm{Ly}\alpha)$ drops to zero \citep{cha93,val93,ver08}. The above predictions do not account for the underlying stellar contribution at \lya. This contribution is not a concern in galaxies dominated by populations of O and early B stars. For reference, the mean \lya~equivalent width of the eight early and mid-O supergiants in \cite{bou12} is $\sim$1 \AA. On the other hand, the correction can be significant for stellar populations dominated by A and late B stars. In addition, the above predictions do not account for the effects of \lya~radiation transfer in the medium surrounding the starburst. Due to its resonant nature, a \lya~photon is absorbed and re-emitted multiple times in \hi. Resonant trapping reduces the mean free path of the \lya~photons increasing significantly their probability of being destroyed by dust, shifted in frequency, or transformed to two-photon emission.  On the other hand, some effects are favorable to the escape of \lya~photons from galaxies. Neutral gas outflows Doppler-shift the \lya~photons from the optically thick line core to the optically thin line wings making it possible for some \lya~photons to escape \citep{kun98,ver06}. \lya~photons may also escape through holes of sufficiently low dust and \hi~column densities \citep{gia96,ate09}. 

In spite of the various \lya~attenuation effects, today, star-forming galaxy candidates have been detected out to redshift $z\sim10$ using various techniques. Galaxies with redshifts of  $z\sim2-5$ are detected via the drop-out technique based on the opacity at the Lyman-limit of the galaxy and the intervening intergalactic medium (IGM; \citealt{gia02,ste03,sha03,ouc04}). Galaxies with $z\gtrsim6$ are detected via the drop-out technique based on the opacity of the IGM at Ly$\alpha$ \citep{bou07,oes10,bou11a,bou11b}. Galaxies with $z\sim2-7$ are also detected using deep narrow band surveys that target the \lya~emission-line  \citep{cow98, hu98, hu04, hu10, rho00, ouc05, ouc08, tan05, ven05, nil07b, nil09, gro07}. Finally, galaxies at the highest redshifts are detected using gravitationally lensing clusters \citep{ota12}. Hundreds of galaxy candidates have been spectroscopically confirmed \citep{ste96, rho03, daw07, iye06,cow10,bra12}, and a few galaxies show impressively high rest-frame equivalent widths of $EW(\rm{Ly}\alpha)\ge$150 \AA~\citep{mal02, shi06, fin08}, which could alternatively be due to the presence of an AGN \citep{cha93}. 

In principle, after removing the observational and astrophysical biases, \lya~can be used for various applications. 1) To probe cosmic star formation rates \citep{kud00, fuj03, gro07, zhe12}. 2) To probe the ionization fraction of the intergalactic medium during the final stages of re-ionization \citep{kas06, mal06, daw07, zhe10, bra12}. 3) To probe the large-scale structure \citep{ouc01, ouc03, ouc04, ouc05, ven02, ven04, wan05, sti05, gaw07}. 4) To identify potential hosts of population III star formation \citep{mal02, shi06, sch07}. 5) And to constrain the nature and evolution of high-redshift galaxies \cite{gia02,nil07a,fin09,sha11,mal12,gon12}.  However, the use of \lya~as a cosmological tool is limited by our understanding of the astrophysical processes that regulate the escape of \lya~photons from galaxies and affect them on their way to telescopes.

The \lya~escape problem can be approached from two complementary angles, doing statistically significant studies of the global properties of the sources and their cosmological context, or studying individual sources in detail. The former is better done in the distant universe, whereas the latter can only be accomplished locally, where the effect of the intergalactic medium is a non-issue, the kinematical and spatial resolutions are higher, and ancillary data is abundant.  The challenges of studying \lya~locally are the necessity of a UV space-based observatory optimized for studying \lya; contamination with the geocoronal \lya~line, which saturates the spectra at very low redshift; and contamination with the Milky Way Galaxy \lya-line trough. Indeed, the interstellar H\,{\sc i} in our Galaxy produces a damped absorption along many sight-lines, which hides the potential \lya~line from star-forming galaxies with redshifts below a few $\times100$ km s$^{-1}$.

There are several observational \lya~studies of galaxies at low-redshift (z$<$0.3). \cite{gia96} studied low resolution spectra of 21 galaxies taken with the \textit{International Ultraviolet Explorer} (\textit{IUE}) and studied the effects of reddening and metallicity on $EW(\rm{Ly}\alpha)$, and Ly$\alpha$/H$\beta$. They found no correlation between these quantities and the UV extinction, the Balmer decrement, or the oxygen abundance. They concluded that rather than the amount of dust, the ISM geometry determines the \lya~escape fraction. Unfortunately, the spectral resolution of \textit{IUE} is insufficient to study the effect of gas outflows on the \lya~escape. \cite{kun98} studied high-dispersion {\it Hubble Space Telescope} (\hst) GHRS spectra of eight nearby \hii~galaxies. They found four emitters, all of which have a velocity offset between the \hi~and the \hii~gas of up to 200 km s$^{-1}$, while the non-emitters show no velocity offset. However, their study suffers from small number statistics. \cite{sca09} analyzed the optical spectra of a sample of 31 $\left<z\right>\sim0.3$ Ly$\alpha$ emitters identified by Deharveng et al. (2008) in low resolution ($\sim8$\,\AA) \textit{Galaxy Evolution Explorer (GALEX)} data. They tried to reproduce the stellar-absorption corrected Ly$\alpha$/H$\alpha$ and H$\alpha$/H$\beta$ ratios using different dust geometries and assumptions for the \lya~scattering in \hi.  They found that the ratios are well reproduced by a clumpy dust distribution, while a uniform dust screen model results in \lya/H$\alpha$~larger than the case B recombination theory value. Unfortunately, the spectral resolution of \textit{GALEX} is insufficient for kinematical studies, and the latter authors were unable to assess the relative importance of dust geometry and outflows.  \cite{ate09} used a 3D \lya~radiation transfer code to model \hst~observations of the very metal-poor dwarf galaxy I Zw 18, which shows \lya~in absorption. Their analysis shows that it is possible to transform a strong \lya~emission of $EW($\lya)\,$\approx$60\,\AA~into a damped absorption, even with low extinction, if the \hi~column density is large. Finally, \cite{ost09} used \hst~to produce \lya, H$\alpha$, and UV continuum maps of six galaxies covering a wide range in luminosity and metallicity, including known \lya~emitters and non-emitters. They found that the bulk of \lya~emerges in a diffuse component resulting from scattering events. They also found the simultaneous presence of \lya~in emission and absorption within spatial scales ranging from a few $\times10$ pc to a few $\times100$ pc. 

Although these studies have addressed main factors affecting the \lya~escape, we still do not understand the relative importance of starburst phase, gas kinematics, gas geometry, and dust extinction, in determining the profile and strength of the \lya~line from star-forming galaxies. We took advantage of the high sensitivity and medium resolution of the Cosmic Origins Spectrograph (COS) aboard \hst~in order to obtain \lya~spectroscopy of a sample of 20 galaxies located a mean redshift of 0.03 and covering a broad range of luminosity, oxygen abundance, and reddening. We used these and ancillary data for establishing correlations between \lya~and fundamental galaxy properties. In section 2 we describe the sample, the observations, and the ancillary data; in section 3 we present our analysis; in section 4 we discuss our results; and in section 5 we summarize and conclude.


\section{SAMPLE, OBSERVATIONS, AND ANCILLARY DATA}

\subsection{Sample}

The galaxies in our sample were H$\alpha$-selected from the Kitt Peak International Spectroscopic Survey data release (KISSR\footnote{The R in ``KISSR'' stands for data with red spectra, i.e., from $6400-7200$ \AA}, \citealt{sal01}). The sample is composed of 12 irregular and eight spiral galaxies distributed in the redshift range $z=0.02-0.06$. At these redshifts, we avoid contamination with the geocoronal and Milky Way (MW) \lya~lines. We cover a factor of three in redshift, which makes it possible to study the effect of distance on the observed properties of \lya. In addition, we cover a broad range in oxygen abundance, $12+\rm{log\,(O/H)}=7.9-9.1$, which makes it possible to  study the effect of metallicity\footnote{In this paper, metallicity is the mass ratio of all elements excluding H and He to all elements.} on the \lya~escape. Metallicity is expected to correlate with the dust content, and dust destroys \lya~photons. We also cover a wide range in H$\alpha$ equivalent width, $EW(\rm{H}\alpha)=27-578$ \AA. The value of  $EW(\rm{H}\alpha)$ is sensitive to the starburst evolutionary phase and to the initial mass function (IMF), e.g., see \cite{lei99}. Finally, we cover over an order of magnitude in \textit{GALEX} \citep{mar05} FUV continuum luminosity, ${\rm log}\,L(1500)=39.0-40.2$ erg s$^{-1}$ \AA$^{-1}$ (uncorrected for reddening). For a given star formation history (SFH), age, and IMF, $L(1500)$ is proportional to the stellar mass in the population dominating the FUV emission of the galaxy. Therefore,  $L(1500)$ can be used to determine the star formation rate of this population (SFR = stellar mass /  time since beginning of star formation). Table~\ref{tab1} summarizes the properties of the sample. The morphological types are based on SDSS images. The H$\alpha$ equivalent widths are from the SDSS spectra when available or from the KISS data otherwise, as indicated in column (10) of Table~\ref{tab1}.


\begin{deluxetable}{llcccccccc}
\tablecolumns{10}
\tablewidth{0pc}
\tabletypesize{\scriptsize}
\tablecaption{Sample Properties}
\tablehead{Galaxy & Morph. & z & d & M$_{\rm B}$ &E(B-V)$_{\rm G}$ & 12+log(O/H) & EW(H$\alpha$) & log\,L$_{1500}$ & Note\\
\hfill & \hfill & \hfill & Mpc & mag & mag & dex& \AA & erg/s/\AA \\
(1) & (2) & (3) & (4) & (5) & (6) & (7) & (8) & (9) & (10)}
\startdata
40 & Irr & 0.026 & 112 & -18.0 & 0.03 & 8.18 & 58 & 39.29 & 1 \\
108 & Irr & 0.024 & 101 & -17.6 & 0.01 & 8.17 & 108 & 39.22 & 1 \\
178$+$ & Irr & 0.057 & 246 & -20.1 & 0.01 & 8.52 & 100 & 40.13 & 1 \\
182 & Irr & 0.022 & 96 & -17.8 & 0.01 & 8.34 & 114 & 39.33 & 1 \\
218 & S & 0.021 & 90 & -19.4 & 0.01 & 8.87 & 66 & 39.65 & 1 \\
242* & Irr & 0.038 & 162 & -19.4 & 0.01 & 8.38 & 435 & 40.25 & 1 \\
271* & Irr & 0.023 & 97 & -17.6 & 0.02 & 8.39 & 29 & 39.13 & 2 \\
298* & S & 0.049 & 210 & -20.0 & 0.02 & 9.07 & 51 & 40.06 & 1 \\
326* & S & 0.028 & 120 & -20.0 & 0.02 & 8.74 & 78 & 39.65 & 2 \\
1084* & S & 0.032 & 137 & -20.5 & 0.05 & 8.79 & 87 & 39.7 & 1 \\
1567* & S & 0.045 & 194 & -20.0 & 0.02 & 8.75 & 29 & 40.18 & 2 \\
1578* & Irr & 0.028 & 120 & -19.6 & 0.02 & 8.14 & 576 & 40.29 & 1 \\
1637 & Irr & 0.035 & 149 & -19.6 & 0.01 & 8.55 & 34 & 39.92 & 1 \\
1785$-$ & Irr & 0.021 & 90 & -17.6 & 0.01 & 7.99 & 199 & 39.07 & 1 \\
1942 & Irr & 0.039 & 169 & -19.3 & 0.03 & 8.59 & 25 & 39.83 & 1 \\
2019 & Irr & 0.034 & 146 & -17.9 & 0.02 & 7.96 & 465 & 39.48 & 1 \\
2021 & S & 0.04 & 171 & -19.6 & 0.02 & 8.79 & 79 & 39.93 & 1 \\
2023 & S & 0.036 & 155 & -19.4 & 0.02 & 9.00 & 58 & 39.77 & 1 \\
2110 & Irr & 0.025 & 108 & -17.3 & 0.01 & 7.86 & 511 & 39.36 & 2 \\
2125 & S & 0.025 & 108 & -18.7 & 0.01 & 8.76 & 85 & 39.31 & 1 \\
Min & \nodata & 0.021 & 90 & -20.5 & 0.01 & 7.86 & 25 & 39.07 & \nodata \\
Max & \nodata & 0.057 & 246 & -17.3 & 0.05 & 9.07 & 576 & 40.29 & \nodata \\
\enddata\\[-15pt]
\tablecomments{(1) KISSR ID. We mark the \lya~emitters with asterisks and the furthest/nearest galaxies with plus/minus signs. The last two rows give the minimum and maximum values in each column. (2) Morphological type. S=spiral. Irr=irregular. (3) Redshift of H$\alpha$ line. (4) Distance computed from the redshift using $d=cz/H_0$, where $c$ is the speed of light, $z$ is the redshift, and we adopt $[\Omega_\lambda,\Omega_M,H_0$]=[0.7, 0.3, 70\,km s$^{-1}$ Mpc$^{-1}$]. (5) Absolute blue magnitude. (6) Color excess based on the Milky Way extinction maps of \cite{sch98} and the reddening law with R$_{\rm{v}}=3.1$ of \cite{fit99}. (7) Oxygen abundance derived as in \cite{sal05}. (8) H$\alpha$ equivalent width. (9) Luminosity at 1500 \AA~from \textit{GALEX}, uncorrected for reddening. (10) Source of the ancillary optical spectrum. 1=SDSS. 2=Same as in Salzer et al. 2005. }
\label{tab1}
\end{deluxetable}


\begin{deluxetable}{lcccc}
\tablecolumns{5}
\tablewidth{0pc}
\tabletypesize{\scriptsize}
\tablecaption{Epochs and Exposure Times of G130M Observations}
\tablehead{Galaxy &Start Time &Start Time & Exp. Time & Exp. Time\\
\hfill &1291 \AA &1318 \AA & (s) 1291 \AA & (s) 1318 \AA\\
(1) & (2) & (3) & (4) & (5)}
\startdata
40 & 2011/06/11 02:07:49 & 2011/06/11 03:26:08 & 1105 & 1105 \\
108 & 2011/07/31 04:47:42 & 2011/07/31 05:09:24 & 1105 & 1105 \\
178 & 2011/06/17 20:55:13 & 2011/06/17 21:16:55 & 1105 & 1105 \\
182 & 2011/07/26 08:13:51 & 2011/07/26 08:35:33 & 1105 & 1105 \\
218 & 2010/02/23 07:33:10 & 2010/02/23 09:09:06 & 1111 & 1111 \\
242 & 2009/12/26 23:56:45 & 2009/12/27 00:18:33 & 1111 & 1111 \\
271 & 2011/05/07 17:41:45 & 2011/05/07 18:03:27 & 1105 & 1105 \\
298 & 2010/01/01 12:41:22 & 2010/01/01 13:03:04 & 1105 & 1105 \\
326 & 2011/04/22 13:36:31 & 2011/04/22 13:58:13 & 1105 & 1105 \\
1084 & 2010/02/05 05:13:19 & 2010/02/05 05:35:01 & 1105 & 1105 \\
1567 & 2010/08/13 05:09:19 & 2010/08/13 05:31:46 & 1150 & 1150 \\
1578 & 2010/02/18 22:07:27 & 2010/02/18 22:29:57 & 1153 & 1153 \\
1637 & 2009/10/24 09:51:17 & 2009/10/24 10:13:44 & 1150 & 1150 \\
1785 & 2011/08/09 10:38:26 & 2011/08/09 11:00:55 & 1142 & 1142 \\
1942 & 2011/06/13 22:46:13 & 2011/06/13 23:08:28 & 1128 & 1128 \\
2019 & 2011/06/19 15:57:20 & 2011/06/19 16:19:49 & 1142 & 1142 \\
2021 & 2009/11/14 00:02:53 & 2009/11/14 00:25:17 & 1147 & 1147 \\
2023 & 2011/06/16 14:29:53 & 2011/06/16 14:52:08 & 1128 & 1128 \\
2110 & 2011/06/16 16:16:21 & 2011/06/16 16:38:50 & 1142 & 1142 \\
2125 & 2011/06/16 17:52:32 & 2011/06/16 18:15:01 & 1142 & 1142 \\
\enddata\\
\label{tab2}
\end{deluxetable}

\subsection{Spectral Class of the Galaxies}

\begin{figure}
\epsscale{1}
\plotone{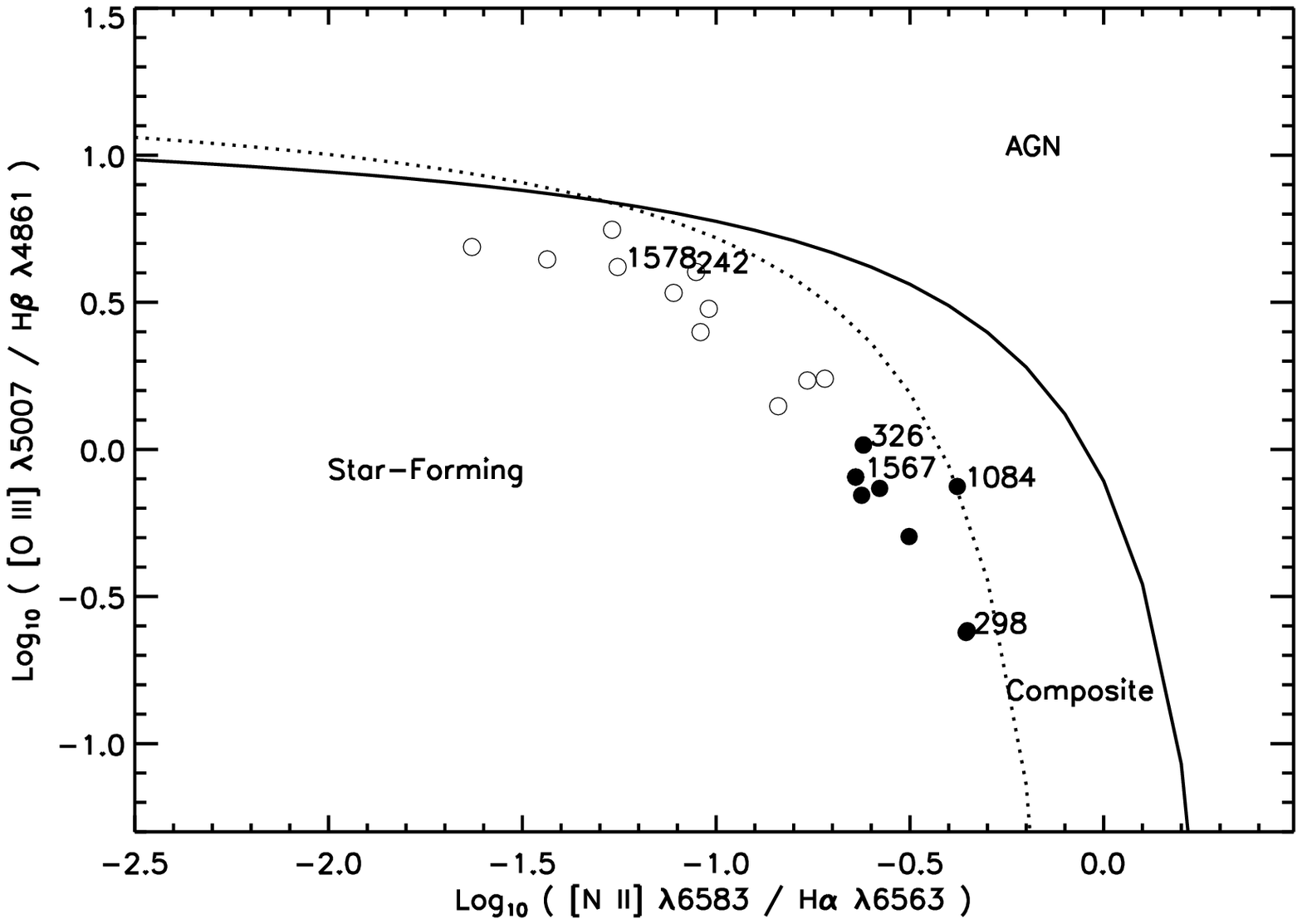}
\caption{Baldwin, Phillips \& Terlevich diagnostic diagram of the spiral (filled symbols) and irregular  (open symbols) galaxies in our sample. We adopt the scheme of \cite{kew06}, where the principal excitation mechanism in the galaxy is star formation if the galaxy lies below the dotted curve, an AGN if the galaxy lies above the solid curve, and a mixture of star formation and Seyfert nucleus, or star formation and LINER, if the galaxy lies between the dotted and solid curves. The dotted curve corresponds to equation 1 in \cite{kau03}. The solid curve corresponds to equation 5 in \cite{kew01}. The symbols of the \lya~emitters are labeled with the KISSR ID of the galaxy.}
\label{fig1}
\end{figure}

We are interested in galaxies where the principal photoionization source is the strong UV radiation from OB stars. This helps in the interpretation of the observed \lya~properties because we do not need to worry about non-stellar ionizing sources such as active galactic nuclei (AGNs), which account for some of the high \lya~EWs at $z>2$ \citep{cha93, zhe10}, and low-ionization nuclear emission-line regions (LINERs), whose nature is not well understood \citep{tan12}. We verified the spectral class of our galaxies using the so-called BPT diagnostic diagram (Baldwin, Phillips \& Terlevich 1981). This was not possible for KISSR 271, for which an optical spectrum was unavailable. As Figure~\ref{fig1} shows, most galaxies are dominated by star formation, but KISSR 1084 may be of composite type. 


\subsection{Observations}

Our observations are part of COS-GTO programs 11522 and 12027 (P.I. J. Green). Descriptions of the COS and its on-orbit performance can be found in \cite{ost11}. We used one orbit with \hst~to observe the brightest FUV knot detected by \textit{GALEX} in each galaxy. We acquired the targets using the ACQ/IMAGE mode and mirror A. Therefore, we obtained two images of each galaxy in the observed-frame wavelength range $1700-3200$ \AA. One after the initial telescope pointing, and the second after the final telescope pointing. We then used the G130M grating to obtain a spectrum of each target in the observed-frame wavelength range $1138-1457$ \AA. We achieved continuous wavelength coverage by using central wavelengths, $\lambda$1291 and $\lambda$1318 at the default focal plane offset position (FP-POS=3). Table~\ref{tab2} lists the epochs and exposure times corresponding to these central wavelengths. The spectra were taken through the circular Primary Science Aperture (PSA), which is 2.5'' in diameter. They contain the aberrated light from objects located up to 2'' or up to $0.9-2.4$ kpc from the aperture's center, depending on the distance to the galaxy. The data were processed with CALCOS v2.13.6 and combined with the custom IDL co-addition routine described in \cite{dan10}. 

\subsection{COS Images}

\begin{figure}
\epsscale{1}
\plotone{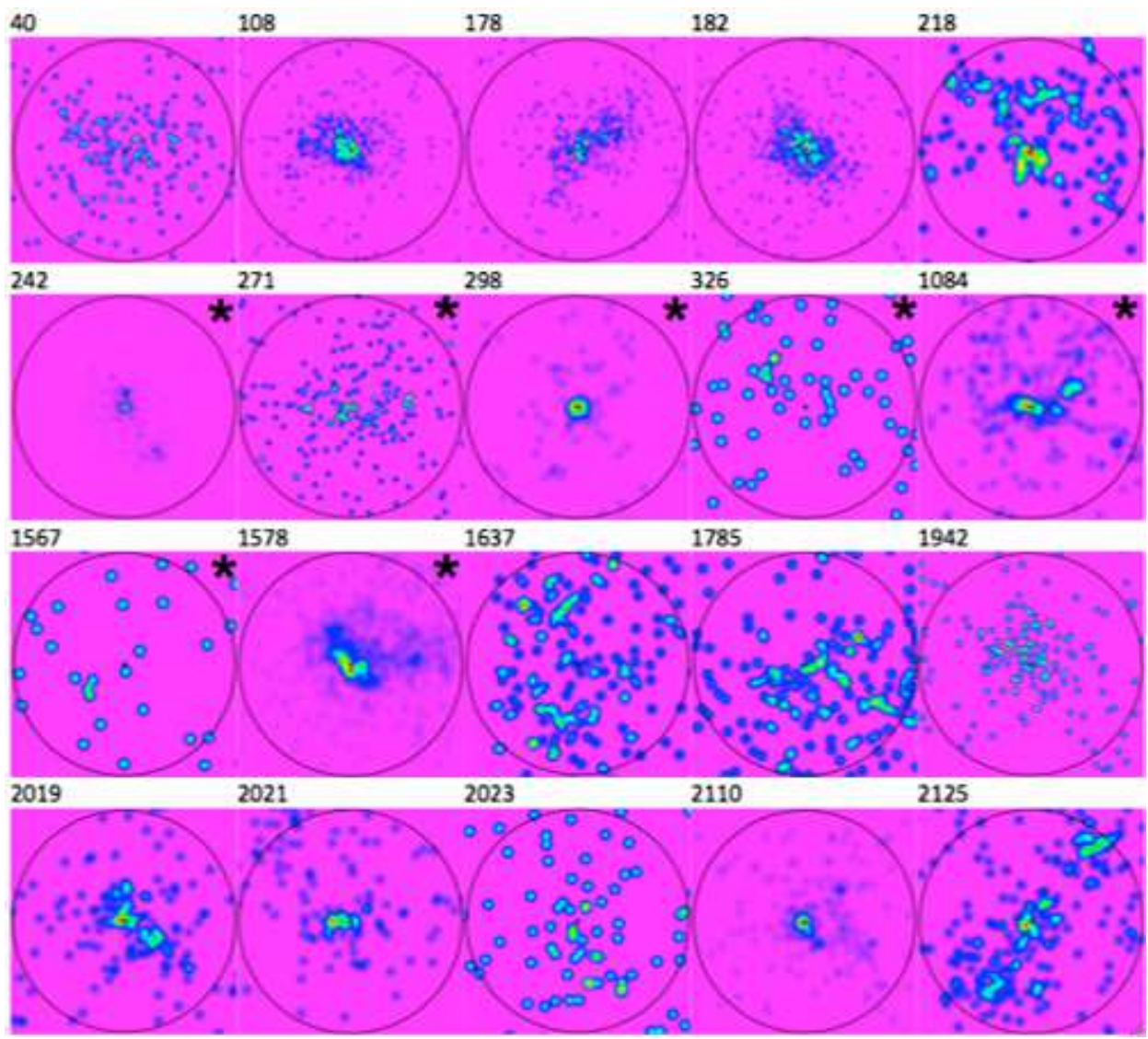}
\caption{Near-ultraviolet ($1700-3200$ \AA) COS TA confirmation images. North is up and east is to the left. The numbers are the galaxy identifiers. The circles represent the 2.5" in diameter COS PSA aperture. The emitters are marked with asterisks. In the rainbow scale, red represents the region of maximum counts, while pink represents regions of low counts.}
\label{fig2}
\end{figure}

Figure~\ref{fig2} shows the NUV target acquisition (TA) confirmation images of the galaxies with the COS aperture overlaid. The center of the aperture is indicated by the small black cross. In 12 cases, a clear source centroid associated with young star clusters is present. 
However, in eight cases, there is no clear centroid of the emission. This is due to the short exposure times of the TA images and in two cases, i.e., for KISSR galaxies 40 and 1567, to TA failures. In spite of the TA failures, spectra were taken. The galaxies that turned out to have net \lya~emission are marked with asterisks. Hereafter, we will call the latter galaxies the emitters. Note that emitters KISSR 271, 326, and 1567, do not show source centroids. Also note that in spite the TA failure, KISSR 1567 turned out to be an emitter. This is in agreement with the finding of \cite{ost09} that the \lya~emission is not necessarily coincident with regions of high FUV surface brightness.
 

\subsection{Ancillary Images}

\begin{figure}
\epsscale{1}
\plotone{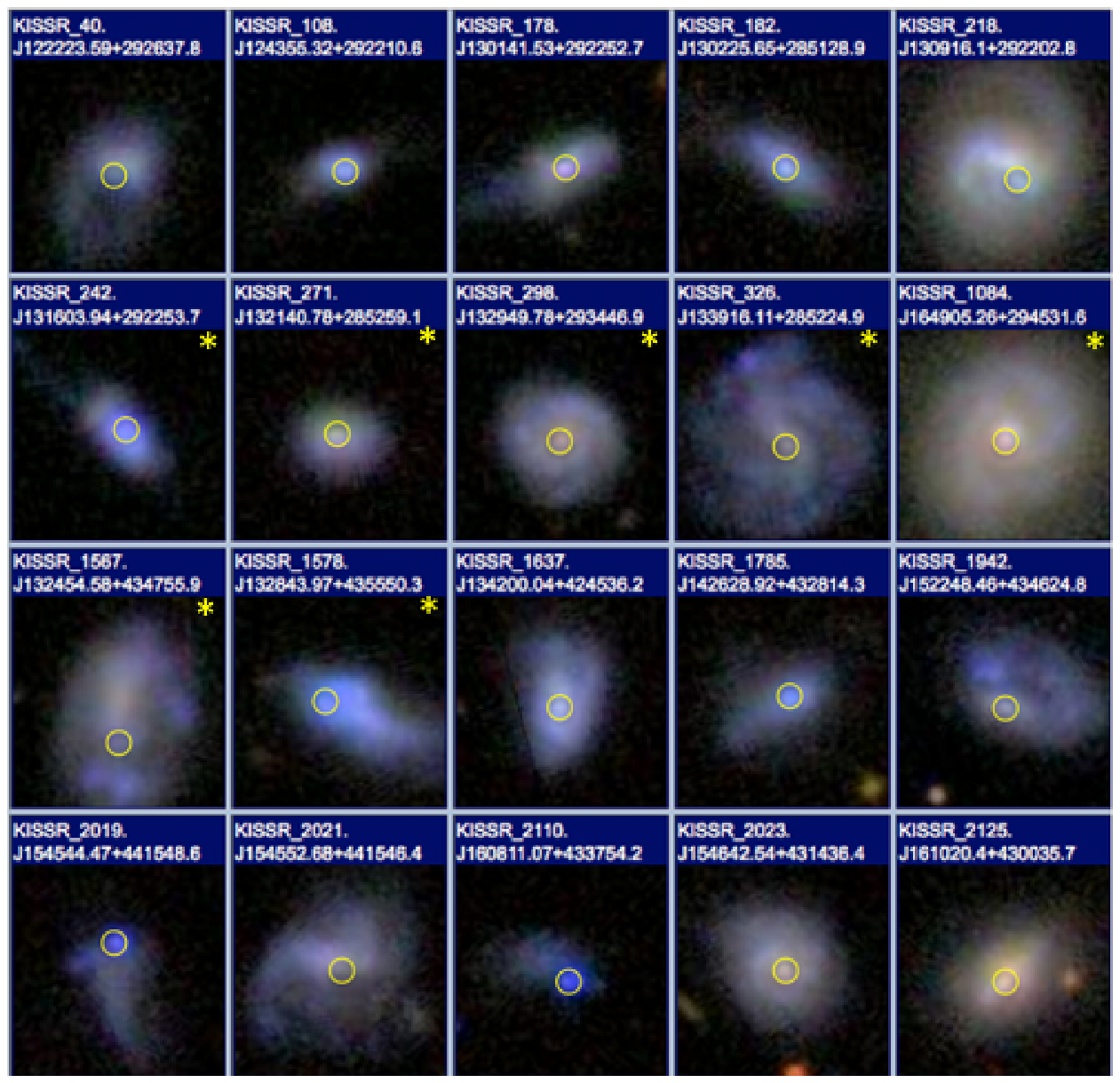}
\caption{Sloan Digital Sky Survey 25"x25" u-, g-, r-, i-, and z-band composite images of the galaxies. The KISSR and SDSS galaxy identifiers are provided. North is up and east is to the left. The \lya~emitters are marked with asterisks. The yellow circles show the COS footprint and pointing.}
\label{fig3}
\end{figure}

We obtained information about the morphology and orientation of the galaxies from archival images. At the time of our analysis, there were no \hst~images available for our targets. The galaxies were imaged by \textit{GALEX} in the FUV at low spatial resolution ($\sim0.5$''). Unfortunately, in these images, the galaxies look like large blobs. However, we found useful Sloan Digital Sky Survey (SDSS, \citealt{yor00}) images of all the targets in our sample. These are shown in Figure~\ref{fig3}, where we have overlaid the COS footprint at the approximate location of the telescope pointing and marked the emitters with asterisks.


Figure~\ref{fig3} shows that most spiral galaxies in our sample have low inclinations except KISSR 2125. While some low-inclination spirals have \lya~in emission, others don't. KISSR 1084, the spiral of composite spectral class is an emitter, and as expected, the highly inclined spiral is a non-emitter. Our two most metal-poor irregulars, KISSR 2019 and 2110, are \lya~absorbers, but irregulars KISSR 242 and 1578 are \lya~emitters. In summary, there are emitters and absorbers among both morphological types. Finally, note that the COS aperture only covers an area of $0.9-2.4$ kpc in radius, depending on the distance to the galaxy. This needs to be considered when comparing with \lya~observations at higher redshifts, since at higher redshift, more of the galaxy is enclosed within the aperture and  \cite{ost09} found that the bulk of the \lya~photons from nearby galaxies are in the diffuse halo.

\subsection{Ancillary Spectra}

We used ancillary optical spectroscopy of the galaxies for determining their redshift, spectral class, metallicity, reddening, SFR, and Wolf-Rayet (WR) star content. Sixteen targets have spectra from the SDSS seventh data release \citep{aba09}. The SDSS spectra are of higher spectral resolution than the KISSR spectra, and they were taken through a circular fiber of $3"$ in diameter that closely matches the size of the COS aperture. The difference between the SDSS and the COS pointings is relevant to the interpretation of the \lya/H$\alpha$ ratio, and to the comparison of the strength of \lya~and the value of $EW$(H$\alpha$). Indeed, the COS aperture encompasses only a portion of each galaxy. Table~\ref{tab3} lists the KISSR identifier (Column~1, the emitters are marked with asterisks), the right ascension  (RA, J2000) and declination (DEC, J2000) of the first \hst~pointing\footnote{The position of the final \hst~pointing is not recorded in the standard output from CALCOS v2.13.6. It is expected to be within 0.3'' of the initial pointing based on results of blind pointing exposures.} (Columns~2 and 3), the RA and DEC of the SDSS spectroscopic pointing (Columns~4 and 5), and the difference between the COS and the SDSS pointings (Columns~6-8). As Table~\ref{tab3} shows, on average, the SDSS pointings are within 0.3" of the COS initial pointings, with a standard deviation of 0.4''. Since the diameter of the COS aperture is 2.5", we do not expect the difference in pointings to constitute a major issue. For KISSR galaxies 326, 1567, and 2110, which do not have SDSS spectra, we were able to recover the optical $2"$-wide slit spectra of \cite{jan05}. Unfortunately, we do not have an optical spectrum for KISSR galaxy 271. 


\begin{deluxetable}{lccccccc}
\tablecolumns{8}
\tablewidth{0pc}
\tabletypesize{\scriptsize}
\tablecaption{Comparison of COS and SDSS Pointings}
\tablehead{Galaxy & COS RA & COS DEC & SDSS RA & SDSS DEC & $\Delta$RA & $\Delta$DEC &  $\Delta_{\rm tot}$\\
\hfill & deg & deg & deg & deg & deg & deg & arcsec\\
(1) & (2) & (3) & (4) & (5) & (6) & (7) & (8)}
\startdata
40 & 185.59833 & 29.44383 & 185.59825 & 29.44394 & 8E-05 & -1E-04 & 0.49 \\
108 & 190.98042 & 29.36964 & 190.98044 & 29.36961 & -2E-05 & 3E-05 & 0.13 \\
178 & 195.42306 & 29.38118 & 195.42304 & 29.38135 & 2E-05 & -2E-04 & 0.62 \\
182 & 195.60688 & 28.85808 & 195.60688 & 28.85807 & -5E-06 & 1E-05 & 0.05 \\
218 & 197.31725 & 29.36739 & 197.31703 & 29.36766 & 2E-04 & -3E-04 & 1.24 \\
242* & 199.01625 & 29.38161 & 199.01632 & 29.38169 & -7E-05 & -8E-05 & 0.38 \\
271* & 200.42000 & 28.88308 & \nodata & \nodata & \nodata & \nodata & \nodata \\
298* & 202.45746 & 29.57972 & 202.45748 & 29.57972 & -2E-05 & 2E-06 & 0.08 \\
326* & 204.81708 & 28.87361 & \nodata & \nodata & \nodata & \nodata & \nodata \\
1084* & 252.27192 & 29.75878 & 252.27195 & 29.75878 & -3E-05 & -2E-06 & 0.12 \\
1567* & 201.22750 & 43.79889 & \nodata & \nodata & \nodata & \nodata & \nodata \\
1578* & 202.18354 & 43.93069 & 202.18357 & 43.93070 & -3E-05 & -6E-06 & 0.10 \\
1637 & 205.50002 & 42.76004 & 205.50038 & 42.76000 & -4E-04 & 4E-05 & 1.30 \\
1785 & 216.62029 & 43.47083 & 216.62029 & 43.47082 & 2E-06 & 1E-05 & 0.05 \\
1942 & 230.70200 & 43.77347 & 230.70197 & 43.77347 & 3E-05 & 2E-06 & 0.11 \\
2019 & 236.43554 & 44.26439 & 236.43552 & 44.26439 & 2E-05 & -1E-06 & 0.08 \\
2021 & 236.46992 & 44.26322 & 236.46991 & 44.26322 & 7E-06 & 2E-06 & 0.03 \\
2023 & 236.67746 & 43.24367 & 236.6774 & 43.24359 & 6E-05 & 8E-05 & 0.35 \\
2110 & 242.04625 & 43.63175 & \nodata & \nodata & \nodata & \nodata & \nodata \\
2125 & 242.58508 & 43.00975 & 242.58507 & 43.00975 & 1E-05 & 0E+00 & 0.05 \\	
\enddata\\
\label{tab3}
\end{deluxetable}


\section{ANALYSIS}

\subsection{Wavelength Zero Point and Effective Spectral Resolution}\label{foreground}

The COS spectra include the intrinsic lines \lya~and a number of low-ionization (LIS) and higher-ionization interstellar lines. We are interested in comparing the offsets of these lines with respect to H$\alpha$. This is for two reasons. First, as mentioned in the introduction, velocity offsets between the \hii~gas traced by H$\alpha$ and the \hi~gas traced by the LIS lines have been showed to favor the escape of \lya~from galaxies. Second, in expanding-shell radiation transfer models the \lya~emission line is redshifted to a few times the shell expansion velocity (e.g., \citealt{ver06}). A similar offset is observed at high redshift when comparing the \lya~redshift to the blueshift of the LIS lines, which give the expansion velocity (e.g. \citealt{sha03}). Thus, offset measurements are useful for establishing how the \lya~strength relates to the interstellar gas kinematics and for constraining radiation transfer models.  The accuracy of the velocity offset measurements depends on the accuracy of the wavelength zero point and on the effective spectral resolution of the data. 

\begin{figure}
\rotate
\epsscale{1}
\plotone{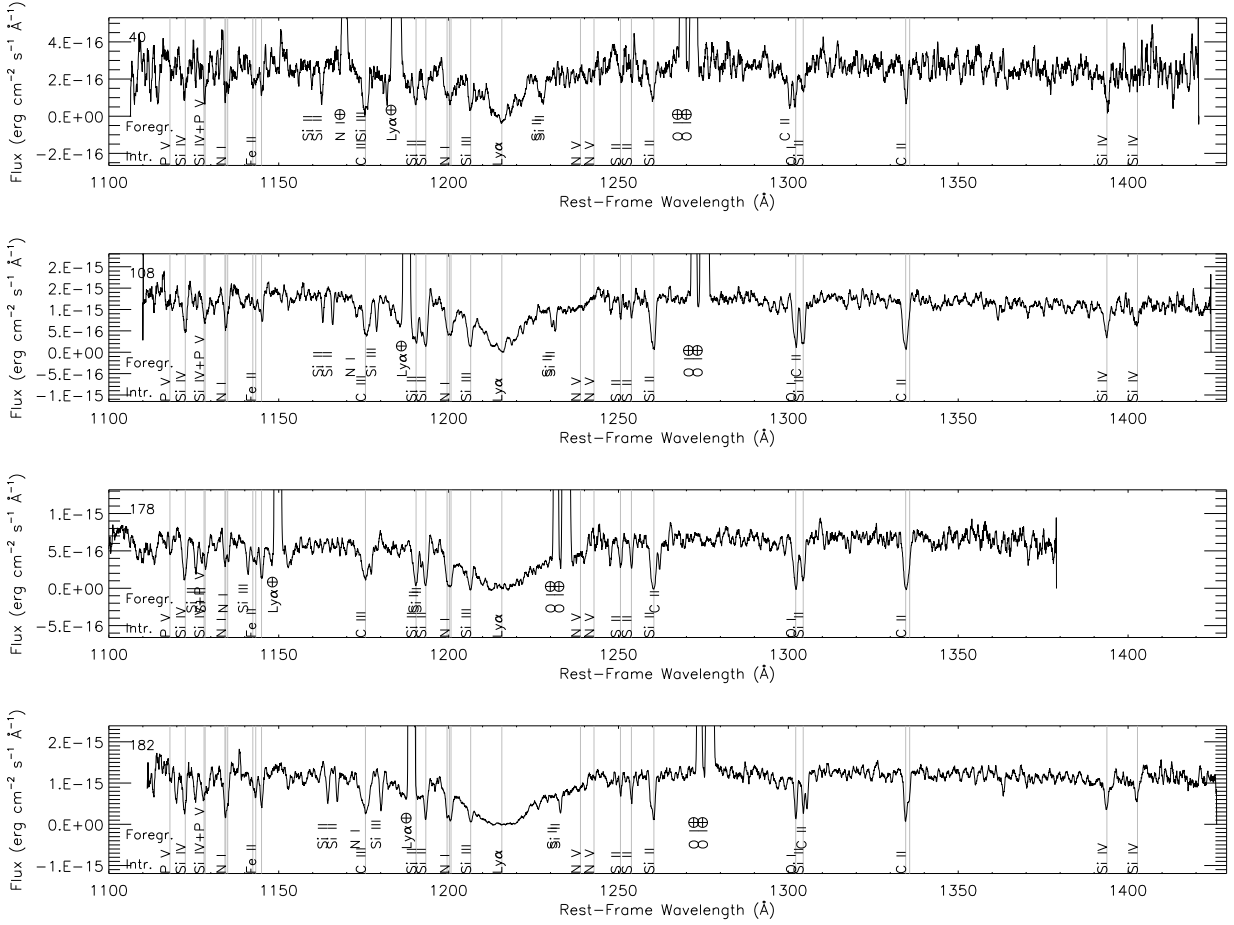}
\caption{COS spectra of galaxies KISSR 40 , 108, 178,  and 182 from top to bottom (KISSR ID provided on the upper left of each panel). The spectra are corrected for the redshift and uncorrected for the reddening. They were first binned to 16  km s$^{-1}$ (nominal resolution of the COS spectra) and then smoothed for clarity. We identify foreground (Foregr.) and intrinsic (Intr.) lines at the bottom of each panel. Geocoronal lines are marked with an encircled plus. The intrinsic lines are marked with vertical gray lines. We did not clip the geocoronal lines in order to show the extent to which they affect the profiles of nearby lines.}
\label{fig4}
\end{figure}

\begin{figure}
\epsscale{1}
\plotone{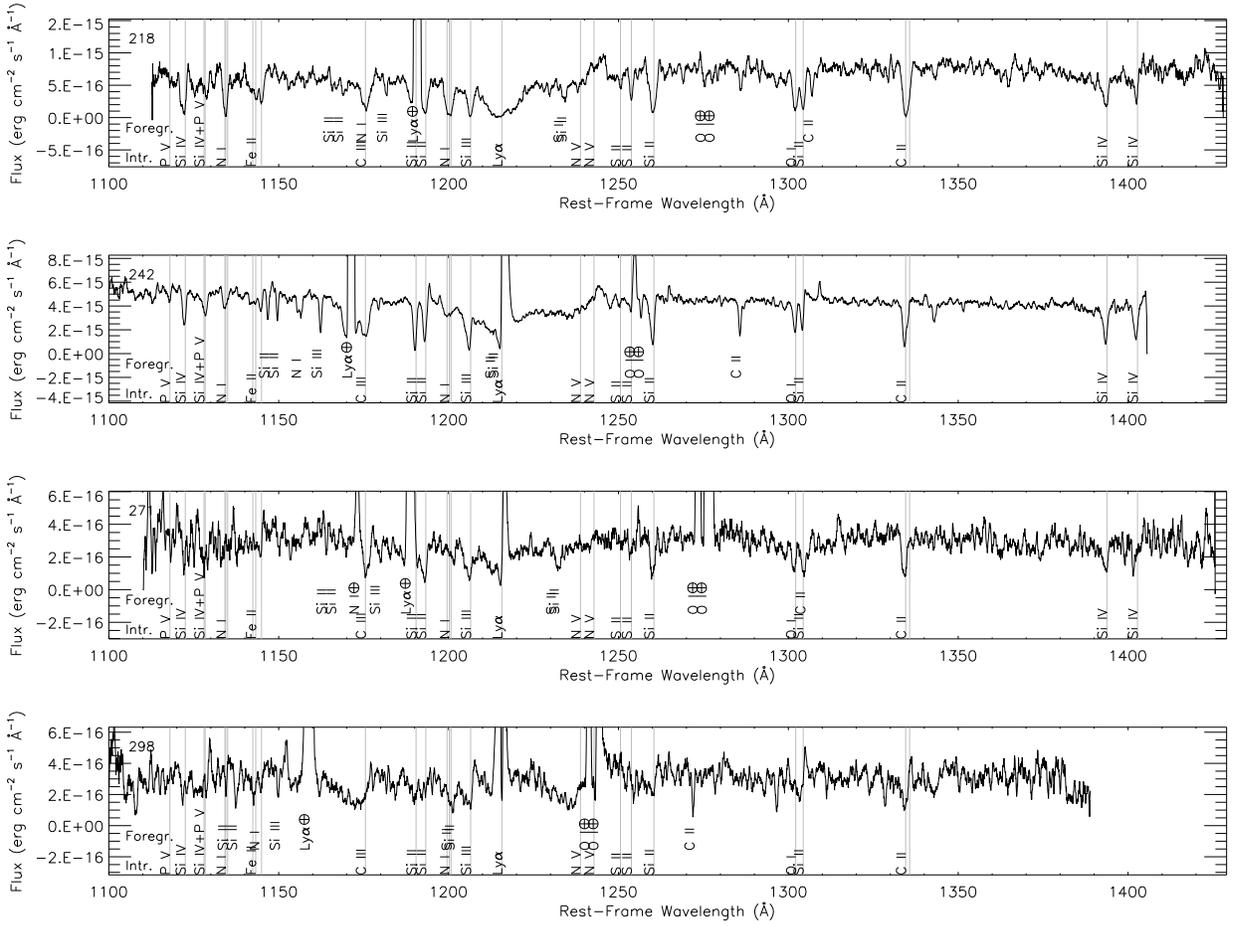}
\caption{Similar to Figure~\ref{fig4} but for galaxies KISSR 218, 242, 271, and 298  from top to bottom.}
\label{fig5}
\end{figure}

\begin{figure}
\epsscale{1}
\plotone{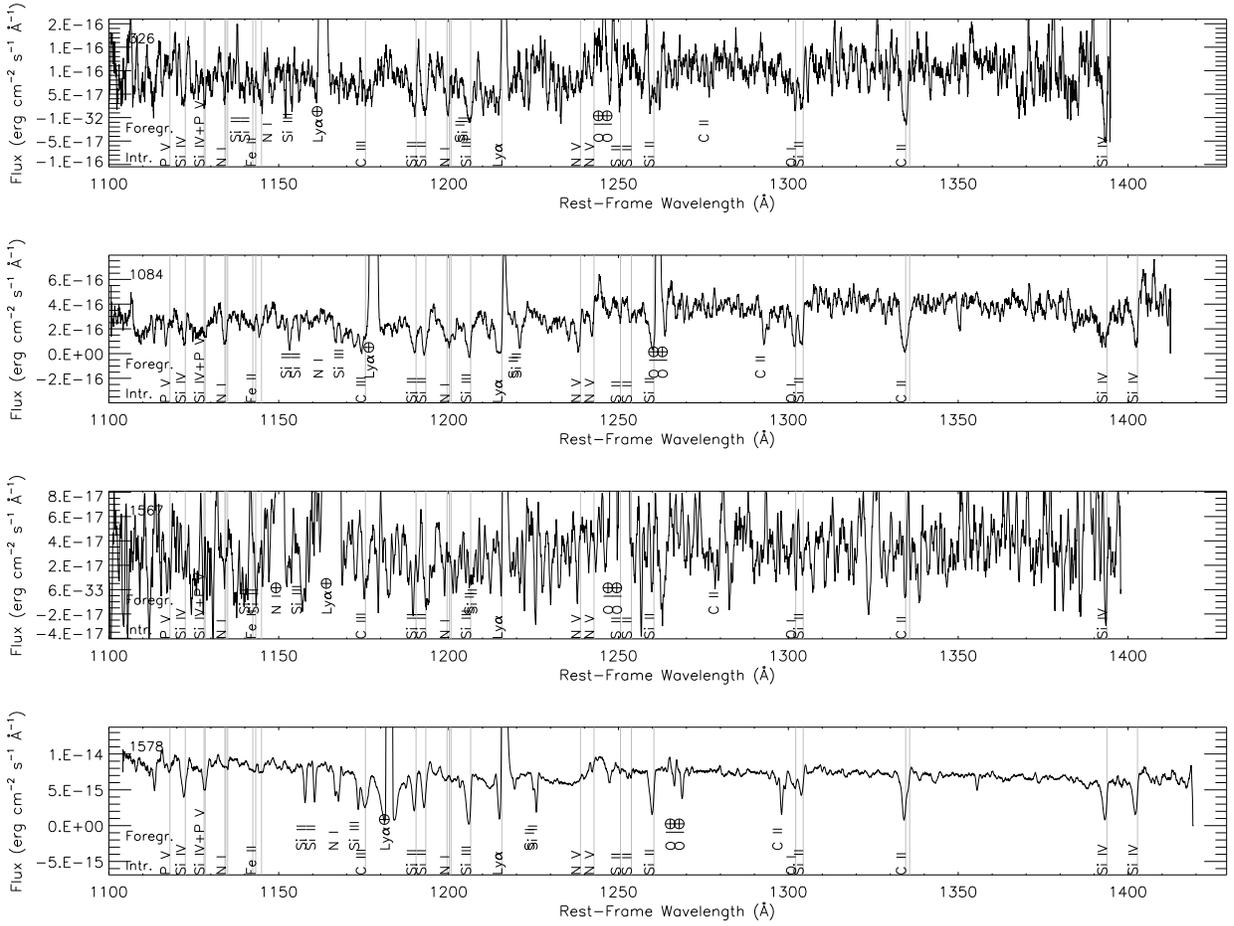}
\caption{Similar to Figure~\ref{fig4} but for galaxies KISSR 326, 1084, 1567, and 1578  from top to bottom.}
\label{fig6}
\end{figure}

\begin{figure}
\epsscale{1}
\plotone{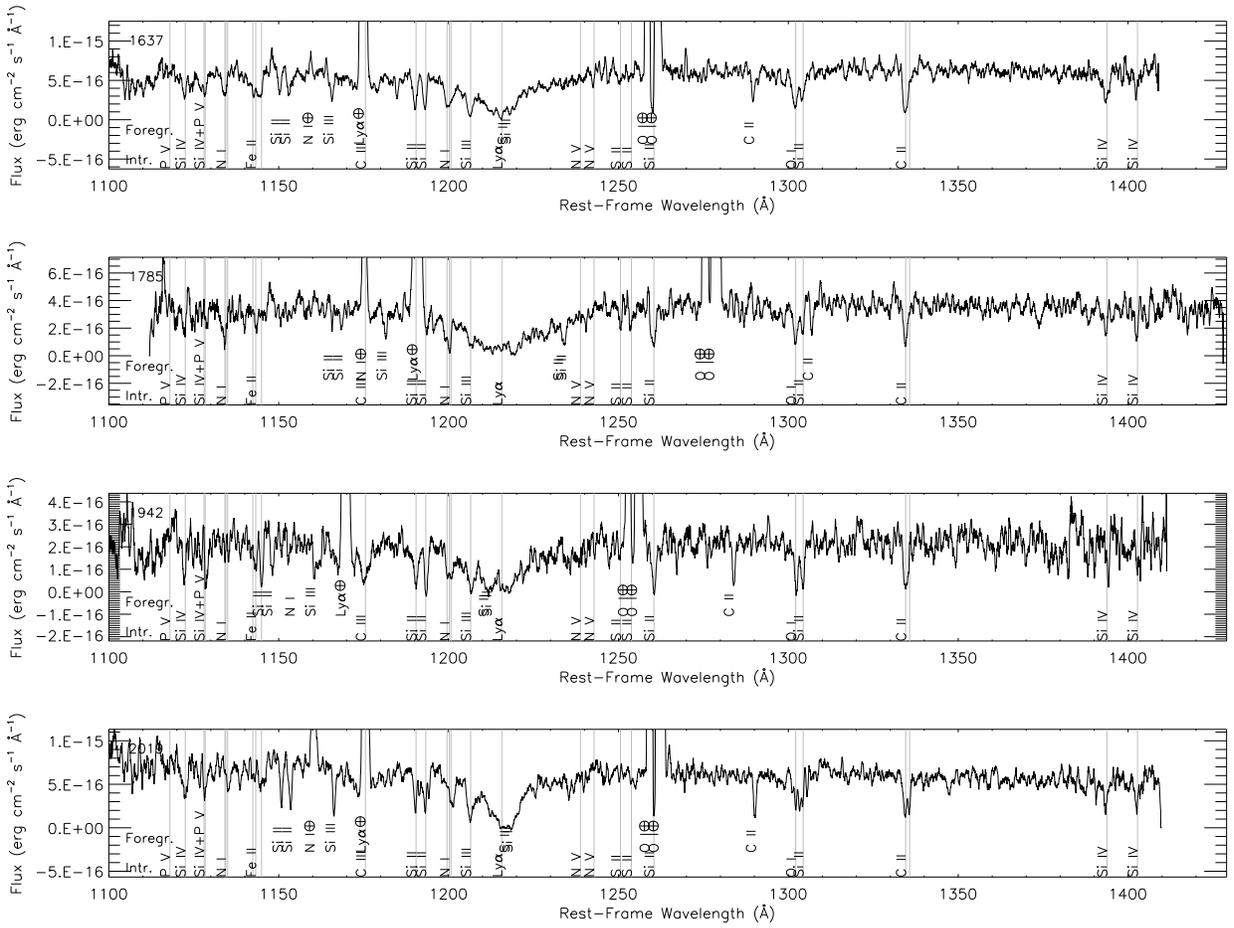}
\caption{Similar to Figure~\ref{fig4} but for galaxies KISSR 1637, 1785, 1942, and 2019  from top to bottom.}
\label{fig7}
\end{figure}

\begin{figure}
\epsscale{1}
\plotone{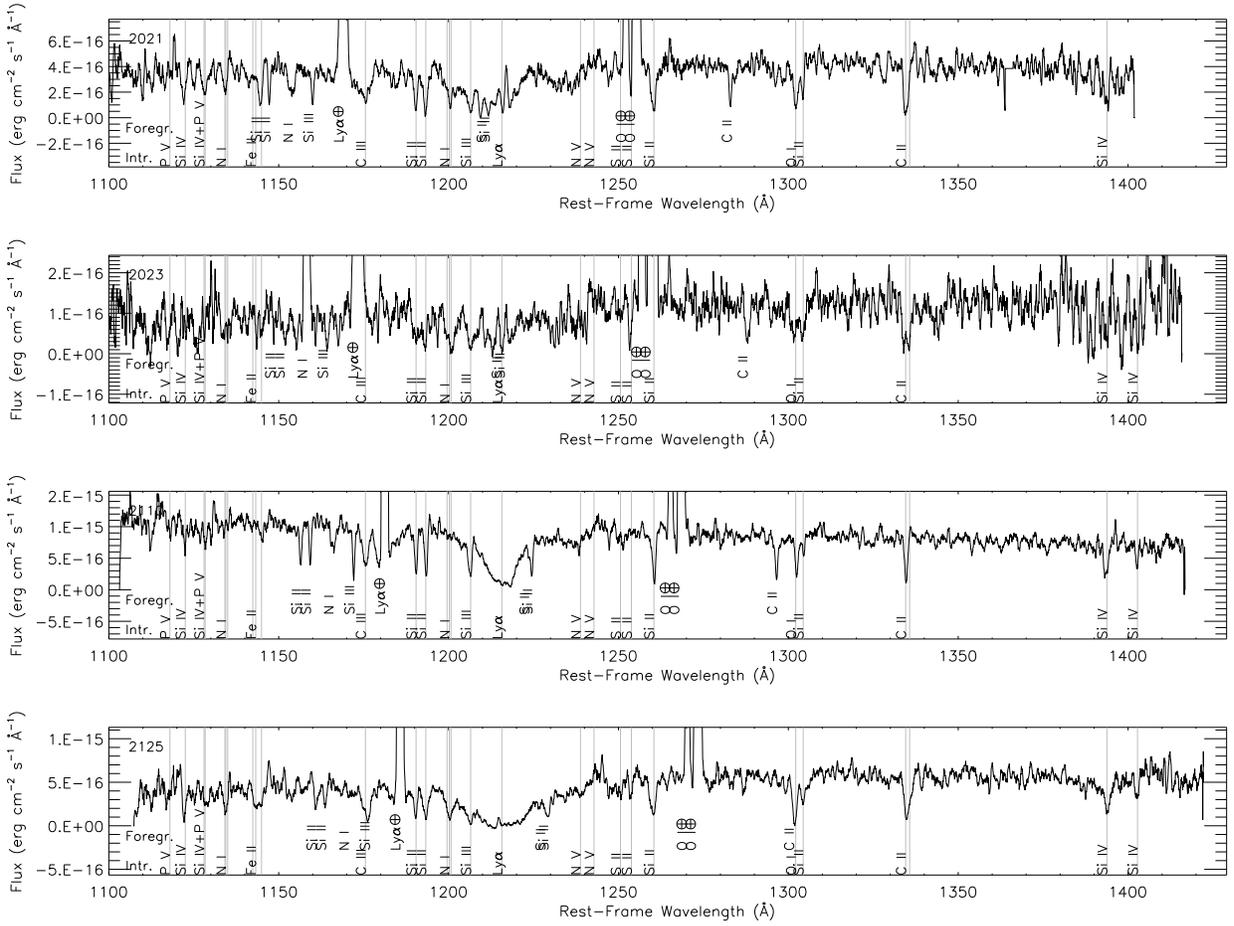}
\caption{Similar to Figure~\ref{fig4} but for galaxies KISSR 2021, 2023, 2110, and 2125  from top to bottom.}
\label{fig8}
\end{figure}

In ultraviolet spectroscopic studies of low-redshift galaxies, it is common to use geocoronal emission lines to check if the wavelength zero point is shifted due to instrumental effects, and Milky-Way metal absorption lines to check if the wavelength zero point is shifted due to the asymmetry of the target in the aperture (e.g., see \citealt{kri11}).  Figures~\ref{fig4}-\ref{fig8} show the COS spectra of the galaxies. They are contaminated with some or all of the geocoronal emission lines \hi~$\lambda$1216, N\,{\sc i}\,$\lambda$1200, and/or O\,{\sc i} $\lambda$1302, 1305; and with the Milky Way (MW) absorption lines, Si\,{\sc ii}~$\lambda$1190, 1193, 1260, 1304, Si\,{\sc iii} $\lambda$1206, O\,{\sc i} 1302, 1305, and/or C\,{\sc ii}~$\lambda$1334, 1336. We measured the centroids of the contaminating lines using a custom routine developed by COS science team member K. France. The routine takes into consideration the COS line spread function. Table~\ref{tab4} lists the mean wavelength offsets of the geocoronal and MW lines. We find that the mean velocity offsets of the geocoronal and MW lines are in the ranges $-33$ to $18$ km s$^{-1}$ and $-1$ to $110$ km s$^{-1}$, respectively.

\begin{figure}
\onecolumn
\rotate
\epsscale{1}
\plotone{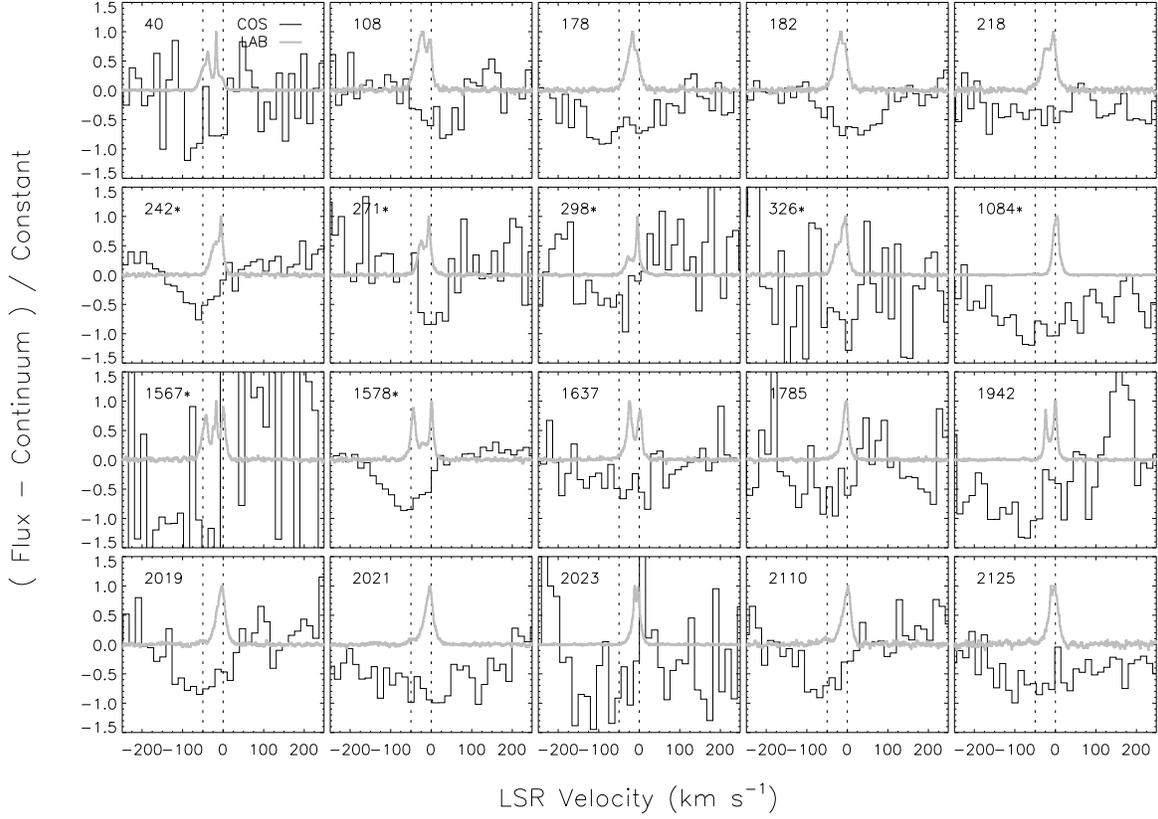}
\caption{IVC \hi~21 cm emission along the lines of sight to our targets (thick gray curves) and MW Si\,{\sc ii}~$\lambda$1190 absorptions (thin black curves). The spectral binnings are the nominal for the LAB and COS data, i.e., 1 km s$^{-1}$ for the former and  16 km s$^{-1}$ for the latter. We give the KISSR ID on the upper left of each panel. The seven galaxies with net \lya~in emission are marked with asterisks. The x-axis gives the velocity in the local standard of rest. The y-axis gives the continuum subtracted flux divided by a constant. The two vertical dotted lines correspond to $v_{\rm LSR}=0$ and 50 km s$^{-1}$.}
\label{fig9}
\end{figure}

The MW lines arise from absorption in halo clouds that may have non-negligible heliocentric velocities. For this reason, we analyzed the H\,{\sc i} 21 cm emission  from intermediate velocity clouds (IVCs)  along the lines of sight to our targets. The data are from the  LAB \hi~survey \citep{kal05}.  In Figure~\ref{fig9} we compare the centroids of the \hi~IVC emission lines and the MW Si\,{\sc ii}~$\lambda$1990 absorptions. The Si\,{\sc ii}~absorptions are corrected for the wavelength offset given by the geocoronal lines. The \hi~21 cm emission is the mean within the radio beam at each velocity. We chose  Si\,{\sc ii}~$\lambda$1990 because it is the only MW line that is not blended with other lines for any of the targets. The \hi~component near 0 km s$^{-1}$ originates in the MW disk. We find that the maximum IVC~cloud velocity is $\sim-50$ km s$^{-1}$ in the local standard of rest frame (e.g., KISSR 40, 1567, and 1578). Although the comparison between the 21 cm emission and the COS data is difficult in cases where the signal to noise of the COS data is low, in general, the 21 cm emission and the Si\,{\sc ii} absorption overlap. However, in some of the highest signal-to-noise cases (e.g., KISSR 242, 1578, 2019, and 2110), the minimum of the absorption is blueward of the bluest 21 cm emission peak. Note that the radio beam is large in comparison with the COS aperture, which makes this wavelength calibration method imperfect. Since large offsets between the 21 cm and the COS data are unlikely to be due to shifts in the wavelength zero point, we decided to only correct the COS data for the small shift in the geocoronal lines. 

Regarding the spectral resolution, the more extended the source, the worse the effective spectral resolution, and the worse the accuracy in the determination of line centroids. Figure~\ref{fig2} shows that some targets are more extended than others. We used the MW lines  Si\,{\sc ii}~$\lambda$1190, 1193, 1260, Si\,{\sc iii} $\lambda$1206, and/or C\,{\sc ii}~$\lambda$1334 for determining the effective spectral resolution of the COS spectra. The dispersion of the G130M grating is 9.97 m\AA/pix and the resolution element is 6 pixels. Therefore, the nominal resolution is 0.06 \AA~or $\sim$16 km s$^{-1}$. The effective spectral resolution of each spectrum is given by the FWHM of the MW absorption lines. For each galaxy, Column 4 of Table~\ref{tab4} lists the mean effective spectral resolution. The mean FWHM of the MW lines is in the range $74-305$ km s$^{-1}$.  We can derive the centroids of the LIS lines with an accuracy of a tenth of the effective spectral resolution, as long as the signal to noise ratio is sufficient.
 

\begin{deluxetable}{llllc}
\tablecolumns{5}
\tablewidth{0pc}
\tabletypesize{\scriptsize}
\tablecaption{Wavelength Zero Point and Effective Spectral Resolution}
\tablehead{Galaxy &
$\left<\rm{v}_{h, \oplus}\right>$&
$\left<\rm{v}_{h, MW}\right>$ &
$\left<\rm{FWHM}_{MW}\right>$ &
$\left<\rm{FWHM}_i\right>$ \\
\hfill &
km s$^{-1}$ &
km s$^{-1}$ &
km s$^{-1}$ &
km s$^{-1}$\\
(1) & (2) & (3) & (4) & (5)}
\startdata
40 & -32 $\pm$ 21  & -78 $\pm$ 27  & 179 $\pm$ 91  & 287   \\
108 & -21 $\pm$ 5  & -1 $\pm$ 5 & 109 $\pm$ 37 & 351   \\
178 & -27 $\pm$ 5  & -62 $\pm$ 29  & 197 $\pm$ 24  & 332   \\
182 & -21 $\pm$ 13  & -30 $\pm$ 12  & 132 $\pm$ 16  & 210   \\
218 & 10  & -26 $\pm$ 51  & 173 $\pm$ 45  & 391   \\
242* & 18 $\pm$ 5 & -34 $\pm$ 8  & 110 $\pm$ 27  & 227   \\
271* & -17 $\pm$ 5  & -10 $\pm$ 43  & 74 $\pm$ 16  & 321   \\
298* & 15  & -49 $\pm$ 17  & 75 $\pm$ 52  & 337   \\
326* & -33 & -90 $\pm$ 60  & 107  & 274   \\
1084* & 10 $\pm$ 2  & -55 $\pm$ 9  & 185 $\pm$ 38  & 446  \\
1567* & -17 $\pm$ 2  & \nodata  & \nodata  & \nodata   \\
1578* & -11  & -70 $\pm$ 4  & 125 $\pm$ 18  & 308   \\
1637 & 2  & -69 $\pm$ 71  & 305 $\pm$ 131 & 359   \\
1785 & -14 $\pm$ 2  & -110 $\pm$ 4 & 142 $\pm$ 10  & 246   \\
1942 & -7 $\pm$ 21  & -95 $\pm$ 61  & 179 $\pm$ 39  & 345   \\
2019 & -16 $\pm$ 6  & -101 $\pm$ 13  & 188 $\pm$ 30  & 238   \\
2021 & -2 $\pm$ 4  & -56 $\pm$ 25  & 174 $\pm$ 81  & 293   \\
2023 & -15 $\pm$ 9  & -38 $\pm$ 72  & 198 $\pm$ 94  & 304   \\
2110 & -8 $\pm$ 15  & -77 $\pm$ 9  & 138 $\pm$ 21  & 186  \\
2125 & -7 $\pm$ 14  & -78 $\pm$ 14  & 274 $\pm$ 74  & 349   \\
Min & -33  & -110 $\pm$ 4  & 74 $\pm$ 16  & 186   \\
Max & 18 $\pm$ 5  & -1 $\pm$ 5  & 305 $\pm$ 131  & 446  \\
\enddata\\[-15pt]
\tablecomments{(1) KISSR ID of the galaxy. We mark the emitters with an asterisk. The last two rows give the minimum and maximum values in each column. In cases where more than one line was measured, we give the standard deviation of the various measurements, as the error. (2) \& (3) Average heliocentric velocities, $v_h$, of geocoronal and MW lines, respectively.  (4) \& (5) Mean FWHM of MW and intrinsic ISM absorption lines. Values of $<110$ km s$^{-1}$ in Column 5 are lower limits.}
\label{tab4}
\end{deluxetable}


\subsection{\lya~and ISM Kinematics}

\begin{figure}
\epsscale{1}
\plotone{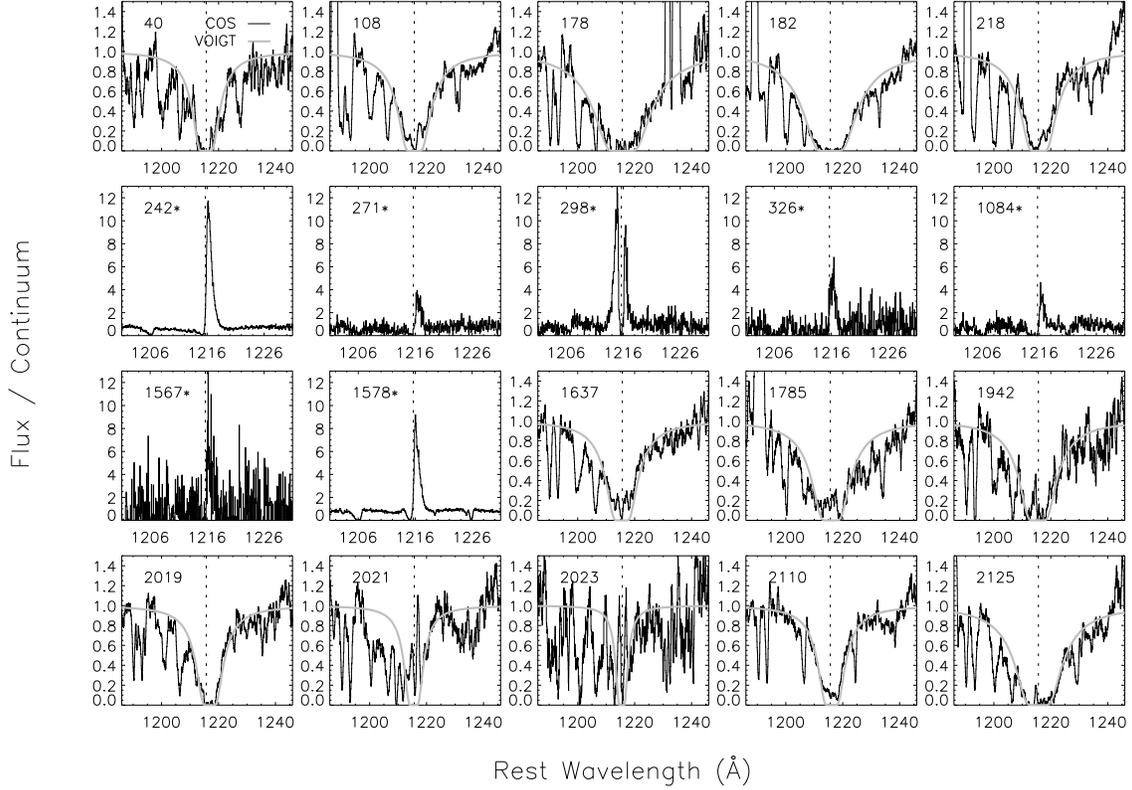}
\caption{Intrinsic \lya~profiles of the galaxies corrected for redshift. We give the KISSR ID on the upper left of each panel. The seven galaxies with net \lya~in emission are marked with asterisks. The spectra are binned to 16 km s$^{-1}$. In addition, the spectra of the absorbers are smoothed for clarity. We overlay the Voigt profiles used for determining the \lya~absorption centroids (gray curves). The vertical dotted lines mark the rest-frame positions of \lya. Note the double \lya~emission and stronger blue peak in KISSR 298. Also note that the \lya~profiles of KISSR 2021 and KISSR 2023 are rather noisy.}
\label{fig10}
\end{figure}

Here we present the velocity offsets of the intrinsic lines mentioned in the previous section. Figure~\ref{fig10} shows the \lya~profiles of the galaxies. Seven objects have net \lya~emission (emitters), eleven objects have damped \lya~absorptions (absorbers), and two objects, KISSR 2021 and 2023 have noisy \lya~profiles blended with MW lines and that appear to be in absorption with perhaps some \lya~emission. We measured the \lya~centroids as follows. For the emitters, the centroid is the point of maximum \lya~flux. For the absorbers, we used the center of the best-fit Voigt profile. The fits to the \lya~absorptions were done by eye using a custom routine written by J. Tumlinson. We overlay the Voigt profiles in Figure~\ref{fig10}.

Excluding KISSR 298, the emitters have P-Cygni like profiles with the emission component redshifted to a mean velocity of $\left<v(\rm{Ly}\alpha)\right>=172\pm$49 km s$^{-1}$. KISSR 298 is a face-on spiral with two \lya~peaks separated by 370 km s$^{-1}$, such that one peak is redshifted and the other is blueshifted and twice as strong. The absorbers have \lya~troughs centered at  a mean velocity of $\left<v(\rm{Ly}\alpha)\right>=0\pm38$ km s$^{-1}$. The \lya~mean velocities exclude KISSR 2021 and 2023 due to their noisy \lya~profiles.

\begin{figure}
\epsscale{1}
\plotone{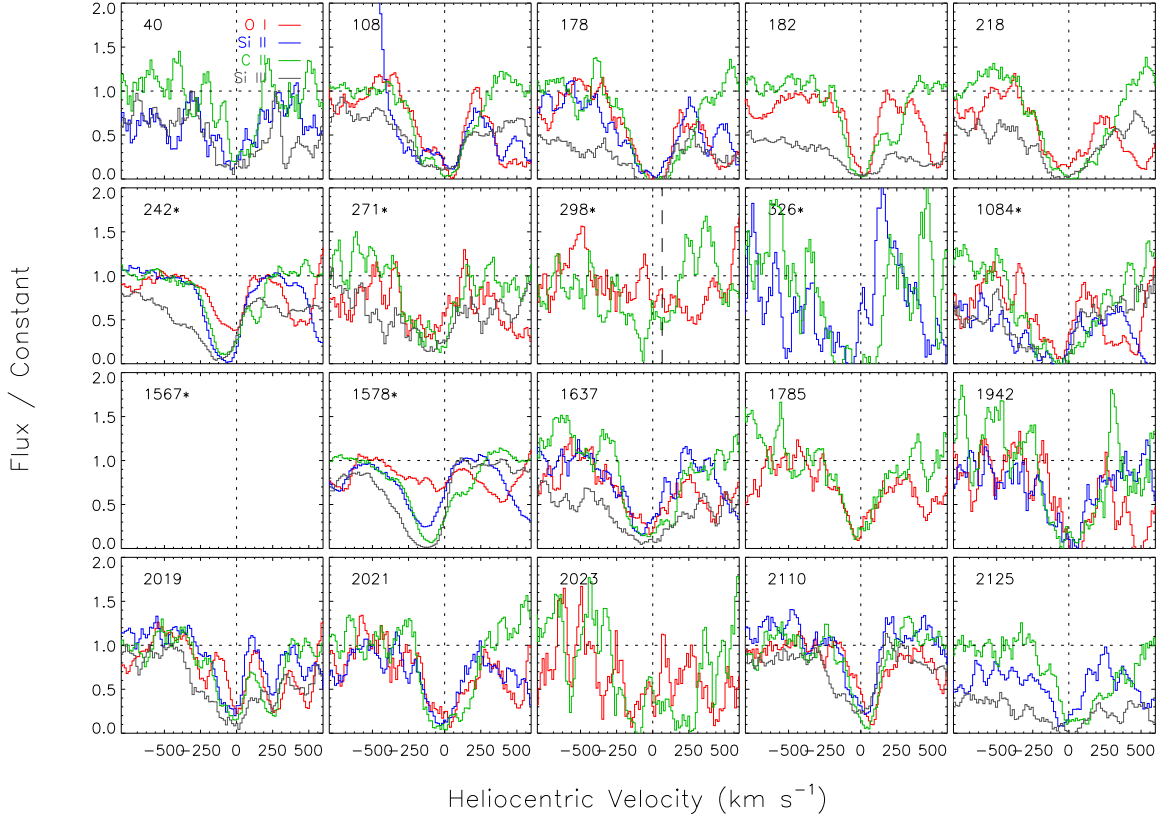}
\caption{Velocity offsets between metal absorption lines intrinsic to the galaxies and H$\alpha$. We include O\,{\sc i} $\lambda$1302 (red), Si\,{\sc ii} $\lambda$1190 (blue), C\,{\sc ii}~$\lambda$1334 (green), and/or Si \,{\sc iii} $\lambda$1206 (gray), except when they are blended with contaminating lines. We give the KISSR ID on the upper left of each panel. The emitters are marked with asterisks. The spectra are binned to 16 km s$^{-1}$ and smoothed for clarity.  The vertical dotted lines mark a velocity offset of zero. For KISSR 298, which is a double-peaked \lya~emitter, we mark the position of the point equidistant from the two \lya~peaks with a vertical dashed line.}
\label{fig11}
\end{figure}

The COS spectra include the \hi~gas tracers O\,{\sc i} $\lambda$1302, Si\,{\sc ii} $\lambda$1190, and C\,{\sc ii}~$\lambda$1334. They also include Si \,{\sc iii} $\lambda$1206, which traces gas in a higher ionization stage. Figure~\ref{fig11} shows the profiles of the previous metal lines in cases where the lines are not severely blended with contaminating lines. We omit emitter KISSR 1567, because for this object the target acquisition failed and the only observable line is \lya. The figure shows that C\,{\sc ii}~$\lambda$1334 and C\,{\sc ii}~$\lambda$1336 are resolved in the highest signal to noise spectra, e.g., for KISSR 242 and 1578.  

For the emitters, the mean blueshifts of O\,{\sc i} $\lambda$1302 and C\,{\sc ii} $\lambda$1334 are $\left<v(\rm{O\,I})\right>=-117\pm47$ km s$^{-1}$ and $\left<v(\rm{C\,II})\right>=-94\pm20$ km s$^{-1}$, respectively. For the absorbers, the corresponding mean velocity offsets are $\left<v(\rm{O\,I})\right>=-23\pm28$ km s$^{-1}$ and $\left<v(\rm{C II})\right>=-21\pm31$, which is almost consistent with a static medium. Note that for spiral KISSR 2023, whose \lya~profile is very noisy, O\,{\sc i} $\lambda$1302 has a large blueshift ($-153$ km s$^{-1}$), such that \lya~emission would be expected. As previously mentioned, there appears to be some \lya~emission in this galaxy which has poor signal to noise. Interestingly, the SDSS image of KISSR 2023 is very similar to that of double-peaked emitter KISSR 298. 

Our kinematical results are shown in Table~\ref{tab5}, which lists the KISSR ID (Column 1); the \lya~profile shape (Column 2); the centroid of the \lya~line (Column 3); the velocity offsets between H$\alpha$ and intrinsic interstellar absorption lines Si\,{\sc ii} $\lambda$1190, Si\,{\sc ii} $\lambda$1193,  Si \,{\sc iii} $\lambda$1206, O\,{\sc i} $\lambda$1302, Si\,{\sc ii} $\lambda$1304, and C\,{\sc ii}~$\lambda$1334 (Columns 4-9); the mean velocity offsets and standard deviations excluding Si \,{\sc iii} $\lambda$1206 (Columns 10 \& 11); the difference between the velocity offset of Si \,{\sc iii} $\lambda$1206 and the value in Column 10 (Column 12);  and for the \lya~absorbers, the \hi~column density of the Voigt profile (Column 13).

In summary, excluding the double emitter KISSR 298, we find that: i) \lya~is in emission and redshifted when high velocity outflows corresponding to expansion velocities of $\sim100$ km s$^{-1}$ are present, ii) \lya~is in absorption and at a velocity close to that of H$\alpha$ when null or low velocity gas flows are present, and iii) the \lya~redshift is higher than the blueshift of the LIS lines by a factor of $1.5-2$. These results are further discussed in section 4. In addition, in cases were the centroid of Si \,{\sc iii} $\lambda$1206 was measured, we find that this line tends to have a blueshift larger than that of the LIS lines by about $30-50$ km s$^{-1}$. 


\begin{deluxetable}{lllllllllllll}
\rotate
\tablecolumns{13}
\tablewidth{0pc}
\tabletypesize{\scriptsize}
\tablecaption{ISM Kinematics}
\tablehead{Galaxy & Ly$\alpha$ & Ly$\alpha$\,1216 & Si\,{\sc ii}\,1190 & Si\,{\sc ii}\,1193 & Si\,{\sc iii}\,1206 & O\,{\sc i}\,1302  & Si\,{\sc ii}\,1304 & C\,{\sc ii}\,1334 & $\left<\Delta v\right>$ & $\sigma(\Delta v)$ & 1206-$\left<\Delta v\right>$ & N(\hi)\\ \hfill & shape & km s$^{-1}$ & km s$^{-1}$ & km s$^{-1}$ & km s$^{-1}$ & km s$^{-1}$ & km s$^{-1}$ & km s$^{-1}$ & km s$^{-1}$ & km s$^{-1}$ & km s$^{-1}$ & cm$^{-2}$\\ 
(1) & (2) & (3) & (4) & (5) & (6) & (7) & (8) & (9) & (10) & (11) & (12) & (13)}
\startdata
40 & Ab & -22 \nodata & \nodata & -29 & \nodata & \nodata & \nodata & -29 & -29 & 0 & \nodata & 5.0E+20 \\
108 & Ab & 0 \nodata & \nodata & -57 & -96 & -48 & \nodata & -50 & -52 & 5 & 44 & 7.5E+20 \\
178 & Ab & 2 \nodata & \nodata & \nodata & \nodata & -35 & -48 & -17 & -33 & 16 & \nodata & 2.3E+21 \\
182 & Ab & -37 \nodata & \nodata & -46 & -47 & -34 & -33 & -47 & -40 & 8 & 7 & 2.0E+21 \\
218 & Ab & 10 \nodata & \nodata & \nodata & \nodata & -32 & -11 & 23 & -7 & 28 & \nodata & 9.0E+20 \\
242 & P-Cygni &  170 $\pm$ 46     & -68 & -73 & -130 & -53 & -104 & -79 & -76 & 19 & 55 & \nodata \\
271 & P-Cygni &  152 $\pm$ 20     & \nodata & \nodata & \nodata & -177 & \nodata & -113 & -145 & 45 & \nodata & \nodata \\
298 & 2 Em &  -118 $\pm$ 11,& \nodata & \nodata & -185 & -162 & -192 & -100 & -151 & 47 & 33 & \nodata \\
\hfill & \hfill &  249 $\pm$ 17 & \hfill & \hfill & \hfill & \hfill & \hfill & \hfill & \hfill & \hfill & \hfill & \hfill\\
326 & P-Cygni &  251 $\pm$ 24     & -162 & -140 & \nodata & -83 & -175 & -88 & -130 & 42 & \nodata & \nodata \\
1084 & P-Cygni &  198 $\pm$ 20     & -169 & -86 & -146 & -109 & -125 & -78 & -113 & 36 & 33 & \nodata \\
1567 & P-Cygni &  105 $\pm$ 41     & \nodata & \nodata & \nodata & \nodata & \nodata & \nodata & \nodata & \nodata & \nodata & \nodata \\
1578 & P-Cygni &  158 $\pm$ 2     & -163 & -147 & -166 & -116 & -149 & -127 & -140 & 19 & 26 & \nodata \\
1637 & Ab & -50 \nodata & -68 & -65 & -60 & -54 & -69 & -9 & -53 & 25 & 6 & 7.0E+20 \\
1785 & Ab & 50 \nodata & \nodata & \nodata & \nodata & -24 & -61 & -42 & -42 & 19 & \nodata & 1.0E+21 \\
1942 & Ab & -50 \nodata & -7 & -2 & \nodata & 6 & -63 & -24 & -18 & 27 & \nodata & 9.5E+20 \\
2019 & Ab & 50 \nodata & -54 & -64 & -37 & -25 & \nodata & -59 & -50 & 18 & -14 & 3.5E+20 \\
2021 & Ab & -50  & -38 & -28 & -19 & -8 & -18 & -14 & -21 & 12 & -2 & 2.5E+20 \\
2023 & Ab & -50  & \nodata & \nodata & \nodata & -153 & -62 & -171 & -128 & 58 & \nodata & 1.0E+20 \\
2110 & Ab & 50 \nodata & -13 & 6 & -76 & 36 & 15 & -2 & 8 & 19 & 85 & 3.5E+20 \\
2125 & Ab & 0  & -56 & -38 & \nodata & \nodata & -44 & 48 & -22 & 48 & \nodata & 1.5E+21 \\
\tablecomments{(1) KISSR ID. (2) Profile shape of \lya. Ab=absorption. P-Cygni=P-Cygni like. 2 Em=Two emission peaks. (3) Velocity offset between the centroid of the \lya~line and H$\alpha$. (4)-(9) Velocity offset between the centroid of the intrinsic metal absorption line and H$\alpha$. (10) Mean velocity offset of intrinsic metal lines excluding Si {\sc iii} $\lambda$1206. (11) Standard deviation of the velocities of the intrinsic metal lines excluding Si {\sc iii} $\lambda$1206. (12) Velocity offset between Si {\sc iii} $\lambda$1206 and the mean of the lower ionization intrinsic metal lines. (13) \hi~column density derived from Voigt profiles.}
\enddata\\[-15pt]
\label{tab5}
\end{deluxetable}

\subsection{$EW(\rm{Ly}\alpha)$, \lya~escape fraction, and $L_{1500}$}

In order to characterize the strength of \lya, we computed the \lya~luminosity $L$(\lya), equivalent width, and escape fraction within the COS aperture. Although the relevant quantity for cosmological studies is the global integrated escape fraction of the galaxy, our measurements are useful for comparing with global values derived from \lya~line images, as they give an idea of the concentration of the emission and of how coincident it is with regions of high FUV surface brightness. Such \lya~maps have been or will be obtained for a portion of our targets. We computed two values of the escape fraction, one using H$\alpha$ and the other using H$\beta$. Both adopt the case B recombination theory predictions of \cite{ost06} for $T_e\sim10^4$ K and $n_e=10^2$ cm$^{-3}$, i.e., \lya/H$\alpha$=8.1 and  \lya/H$\beta$=23.1. $L$(\lya). The escape fractions were computed before and after correcting for the dust attenuation.  We also computed the COS luminosity of the continuum at 1500 \AA, $L_{1500}$, for comparison with the value from \textit{GALEX}, which is given in Table~\ref{tab1}. In order to obtain $L_{1500}$, we extrapolated beyond the red edge of our data which is less than 1500 \AA. In this paper, $L_{1500}$ is used for deriving the mass of the dominant stellar population and the star formation rate (see below). 

For correcting the hydrogen line fluxes and $L_{1500}$~for the dust attenuation, we derived the total color excess, $E(B-V)_t$, from the observed H$\alpha$/H$\beta$ line intensity ratio, after correcting the Balmer lines for the underlying stellar absorption. The stellar absorption correction was done by adding 2 \AA~to the  equivalent widths of the Balmer lines, following \citep{mac85}. This increases the H$\alpha$ flux by 2\% and the H$\beta$ flux by 13\% (median values for the sample).  We did not correct the interstellar \lya~for the stellar contribution at \lya, since as mentioned in the introduction, this correction is expected to be small for stellar populations dominated by O and early B stars, as is the case for the \lya~emitters in our sample. That O and early B stars dominate for the emitters is shown in section 3.5. The \lya~emission line fluxes were obtained by numerical integration. The Balmer line fluxes were  obtained from the SDSS data when available and from the KISS data otherwise. The SDSS spectra have higher spectral resolution.

The total color excess is given by $E(B-V)_t=E(B-V)_G+E(B-V)_i$, where $E(B-V)_G$ and $E(B-V)_i$ are the Galactic and intrinsic contributions, respectively. According to the Milky Way reddening maps of \cite{sch98}, $E(B-V)_G$ is small towards our targets (see Table~\ref{tab1}), thus most of the extinction must be intrinsic. Since \cite{val93} and \cite{cal92} showed that by using a metallicity-dependent reddening law one could recover the theoretical \lya/H$\beta$ ratio, we tried using a MW  law for the spiral galaxies and an SMC  law for the irregular galaxies. For this purpose, we used the reddening laws tabulated in \cite{ost06}. However, we found that the $E(B-V)_t$ values derived from these two laws are very similar. In the end, and as in \cite{cal00}, we adopted a MW law for deriving $E(B-V)_t$. For the theoretical H$\alpha$/H$\beta$ ratio we used 2.86, which is the case B recombination theory prediction for $T_e=10^4$\,K and $n_e=100$ cm s$^{-3}$ of \cite{ost06}.  

In order to see if the relation $E(B-V)_s=0.44\times E(B-V)_t$ of \cite{cal00} holds for our data, we computed the color excess of the stellar continuum, $E(B-V)_s$, using the slope of the FUV continuum, $\beta$. For obtaining $E(B-V)_s$ we used the method outlined in \cite{wof11}, where the slope of the Milky-Way de-reddened continuum is compared to that of a dust-free model of the appropriate metallicity. Following \cite{cal00}, $E(B-V)_s$ was derived using the starburst reddening law. The comparison was done over the wavelength range from 1200 \AA~to the red edge of the data. The red edge is redshift-dependent, as shown in Figures 4-8, and it is always less than 1500 \AA. This range is different from what is used in \cite{cal00}, i.e.,  $1250-1950$ \AA.

Our results are shown in Table~\ref{tab6}, which gives the KISSR ID of the galaxy (Column 1); the rest-frame value of $EW($\lya) (Column 2); $L$(\lya), $L_{1500}$ from the COS spectra, $L$(H$\alpha$), and $L$(H$\beta$) (Columns 3-6, uncorrected for dust extinction); the escape fractions within the COS aperture, i.e. \lya/H$\alpha$/8.1$\times$100 and \lya/H$\beta$/23.1$\times$100 (Columns 7 \& 8); $E(B-V)_t$ from the Balmer lines and $E(B-V)_s$ from the slope of the FUV continuum (Columns 9 \& 10); and the reddening-corrected values of $L_{1500}$, H$\alpha$/H$\beta$, and the escape fractions (Columns 11-14). 

In summary, we find that $E(B-V)_s=1.27\times E(B-V)_t$ (adopting the median of the sample), i.e., that the stellar continuum is more reddened than the nebular continuum, contrary to what was found by \cite{cal00}. The different result is attributed to the larger optical and ultraviolet apertures used by \cite{cal00}, and to the different wavelength range used by the latter authors for fitting the UV continuum. We also find that the rest-frame $EW$s of the emitters are in the range $1-12$\,\AA, and that the reddening corrected escape fractions of the emitters are in the range $1-12$\%. The latter two results are further discussed in section 4. Finally, we found that $L_{1500}$ from the COS data is lower by a factor of a few compared to the value from \textit{GALEX} (compare column 4 of Table~\ref{tab6} with column 9 of Table~\ref{tab1}). This is attributed to the larger aperture of \textit{GALEX}. 


\begin{deluxetable}{lccccccccccccc}
\tablecolumns{14}
\rotate
\tablewidth{0pc}
\tabletypesize{\scriptsize}
\tablecaption{Ly$\alpha$ Escape Fractions}
\tablehead{Galaxy & 
EW(\lya) & 
log$L$(\lya) & 
log$L_{1500}$ & 
log$L$(H$\alpha$) & 
log$L$(H$\beta$) & 
$f$(H$\alpha$) & 
$f$(H$\beta$) & 
E(B-V)$_{t}$ &
E(B-V)$_{s}$ & 
log$L_{1500}$ &
H$\alpha$/H$\beta$ &
$f$(H$\alpha$) & 
$f$(H$\beta$) \\
\hfill & \AA & erg s$^{-1}$ & erg s$^{-1}$ & erg s$^{-1}$ & erg s$^{-1}$ & \% & \% & mag & mag & erg s$^{-1}$ & \hfill & \% & \%\\
\hfill & \hfill & uncorr & uncorr & uncorr & uncorr & uncorr & uncorr & Balmer & $\beta$ & corr & \hfill & corr & corr \\
(1) & (2) & (3) & (4) & (5) & (6) & (7) & (8) & (9) & (10) & (11) & (12) & (13) & (14)}
\startdata
40	&	$<0$	&	\nodata	&	38.6	&	40.02	&	39.54	&	\nodata	&	\nodata	&	0.04	&	0.21	&	39.5	&	3.1	&	\nodata	&	\nodata	\\
108	&	$<0$	&	\nodata	&	39.1	&	40.38	&	39.85	&	\nodata	&	\nodata	&	0.15	&	0.17	&	39.9	&	3.5	&	\nodata	&	\nodata	\\
178	&	$<0$	&	\nodata	&	39.6	&	41.21	&	40.64	&	\nodata	&	\nodata	&	0.23	&	0.20	&	40.5	&	4	&	\nodata	&	\nodata	\\
182	&	$<0$	&	\nodata	&	39.1	&	40.41	&	39.88	&	\nodata	&	\nodata	&	0.16	&	0.19	&	39.9	&	3.6	&	\nodata	&	\nodata	\\
218	&	$<0$	&	\nodata	&	38.9	&	40.49	&	39.87	&	\nodata	&	\nodata	&	0.33	&	0.41	&	40.6	&	4.6	&	\nodata	&	\nodata	\\
242*	&	10	&	41.20	&	40.1	&	41.61	&	41.07	&	5	&	6	&	0.18	&	0.21	&	41.0	&	3.7	&	12	&	14	\\
271*	&	1	&	38.74	&	38.5	&	\nodata	&	\nodata	&	\nodata	&	\nodata	&	0.17	&	0.28	&	39.7	&	\nodata	&	\nodata	&	\nodata	\\
298*	&	12	&	40.16	&	39.3	&	40.65	&	40.06	&	6	&	2	&	0.28	&	0.39	&	41.0	&	4.3	&	12	&	15	\\
326*	&	4	&	39.30	&	38.7	&	40.22	&	39.61	&	2	&	0	&	0.16	&	0.38	&	40.3	&	4.4	&	7	&	9	\\
1084*	&	1	&	38.99	&	39.0	&	41.03	&	40.37	&	0	&	1	&	0.42	&	0.44	&	41.0	&	5.2	&	1	&	1	\\
1567*	&	3	&	38.71	&	38.3	&	40.25	&	39.63	&	0	&	1	&	0.15	&	0.31	&	39.6	&	4.7	&	2	&	3	\\
1578*	&	7	&	41.00	&	40.0	&	41.55	&	41.04	&	3	&	4	&	0.12	&	0.10	&	40.4	&	3.4	&	6	&	7	\\
1637	&	$<0$	&	\nodata	&	39.2	&	40.43	&	39.93	&	\nodata	&	\nodata	&	0.10	&	0.30	&	40.5	&	3.3	&	\nodata	&	\nodata	\\
1785	&	$<0$	&	\nodata	&	38.5	&	40.38	&	39.83	&	\nodata	&	\nodata	&	0.18	&	0.19	&	39.3	&	3.7	&	\nodata	&	\nodata	\\
1942	&	$<0$	&	\nodata	&	38.9	&	40.00	&	39.44	&	\nodata	&	\nodata	&	0.21	&	0.28	&	40.1	&	3.9	&	\nodata	&	\nodata	\\
2019	&	$<0$	&	\nodata	&	39.1	&	40.63	&	40.17	&	\nodata	&	\nodata	&	0.01	&	0.14	&	39.7	&	2.9	&	\nodata	&	\nodata	\\
2021	&	$<0$	&	\nodata	&	39.2	&	40.59	&	40.03	&	\nodata	&	\nodata	&	0.21	&	0.30	&	40.5	&	3.8	&	\nodata	&	\nodata	\\
2023	&	\nodata	&	\nodata	&	38.6	&	40.58	&	39.91	&	\nodata	&	\nodata	&	0.45	&	0.31	&	39.9	&	5.5	&	\nodata	&	\nodata	\\
2110	&	$<0$	&	\nodata	&	39.1	&	40.33	&	39.86	&	\nodata	&	\nodata	&	0.00	&	0.09	&	39.5	&	3	&	\nodata	&	\nodata	\\
2125	&	$<0$	&	\nodata	&	38.9	&	40.73	&	40.03	&	\nodata	&	\nodata	&	0.49	&	0.30	&	40.2	&	5.7	&	\nodata	&	\nodata	\\
Min	&	1	&	38.71	&	38.3	&	40.00	&	39.44	&	0	&	0	&	0.00	&	0.09	&	39.3	&	2.9	&	1	&	1	\\
Max	&	12	&	41.20	&	40.1	&	41.61	&	41.07	&	6	&	6	&	0.49	&	0.44	&	41.0	&	5.72	&	12	&	15	\\
\enddata\\[-15pt]
\label{tab6}
\end{deluxetable}

\subsection{SED Models}\label{sec4_models}

We studied how the strength of \lya~relates to the starburst phase and star formation rate of the dominant stellar population by comparing the COS observations with synthetic dust-free spectra computed with the widely used package Starburst99 (S99, \citealt{lei99, lei10, vaz05}). The observed star formation rate is that within the COS aperture. The value of $EW($\lya) depends on the metallicity, IMF, star formation history, and evolutionary phase of the stellar population (see models by \citealt{val93, cha93, ver08}).  Because empirical spectral stellar libraries in the UV are only available at the Galactic ($Z_\odot=0.013$, \citealt{asp09}) and LMC/SMC ($Z_{\rm{LMC}}=0.007$ and $Z_{\rm{SMC}}=0.002$, \citealt{mae99}) metallicities, we have to rely on theoretical libraries at other metallicities. We used the theoretical libraries of \cite{lei10} corresponding to $Z=0.001$, 0.004, 0.008, 0.020, and 0.040. For comparison, we also computed models with the empirical Galactic and LMC/SMC stellar libraries. We will refer to simulations based on the theoretical and empirical libraries as the theoretical and empirical models, respectively. The theoretical models use stellar evolution tracks corresponding to $Z=0.020$ and model atmospheres corresponding to $Z=Z_\odot$. The empirical models use tracks corresponding to $Z=0.020$ or $Z=0.004$. Our grid of models includes single stellar population (SSP) and continuous star formation (CSF) scenarios. Continuous star formation is more appropriate in our case, as the observed spectra include light from regions of a few kpc in size and such large regions must contain UV-bright star clusters that span a range of ages. The SSP models are for comparison.  We adopted a Kroupa IMF \citep{kro01}, as at the high-mass end, it is considered to be universally applicable \citep{bas10}. We used the Geneva stellar evolution tracks for non-rotating single-stars with high-mass loss \citep{sch92, mey94} because models that account for stellar rotation and binarity \citep{eld08, lev12} are still under development and require calibration against observations over our range of metallicities.  Finally, we did not compute models for times earlier than 2 Myr because this is the age of the youngest LMC main-sequence O stars, or greater than 30 Myr after the beginning of star formation since the SSP models are affected by incompleteness of the stellar libraries beyond this time and the continuum slope and age-sensitive stellar-wind lines of CSF models do not significantly change beyond this time.

\subsection{Starburst Phase, Stellar Mass, and Star Formation Rate}

The age of an FUV-bright stellar population can be derived spectroscopically by comparing the observed profiles of age sensitive lines originating in the winds of massive stars with model spectra. The strong N\,{\sc v}~$\lambda$1238.8, 1242.8, Si\,{\sc iv}~$\lambda$1393.8, 1402.8, and C\,{\sc iv}~$\lambda$1548.2, 1550.8 resonant doublets are widely used for this purpose \citep{tre01, cha03, wof11}. In summary, the P-Cygni profiles of the N\,{\sc v} and the C\,{\sc iv} doublets decrease in strength as the O stars evolve and expire. On the other hand, the Si\,{\sc iv}~doublet develops a P-Cygni profile in giant and supergiant O stars. Unfortunately, the C\,{\sc iv} doublet and sometimes the Si\,{\sc iv} doublet fall outside of the wavelength range that we observed with COS. In addition, the N\,{\sc v} doublet is sometimes contaminated with geocoronal emission. However, for targets showing clean and strong N\,{\sc v} P-Cygni profiles, the age of the dominant population is younger than $\sim$10 Myr (see figure 5 in \citealt{wof11}). Furthermore, a large H$\alpha$ equivalent also indicates the presence of a young population \citep{ken98, lei99}.

\begin{figure}
\epsscale{1}
\plotone{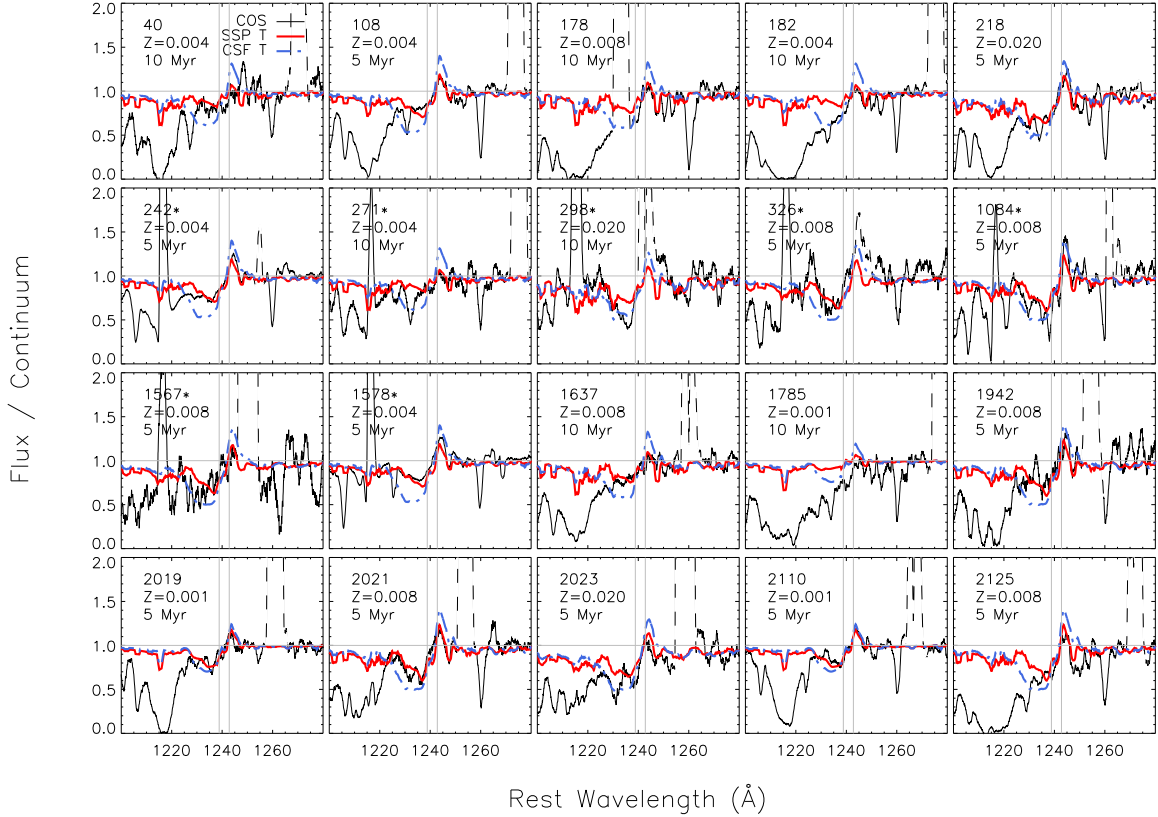}
\caption{Comparison of observed and computed N\,{\sc v}~1240 resonance doublets. The vertical gray lines give the rest wavelengths of the N\,{\sc v} lines. The observations (thin black curves) are sometimes contaminated with geocoronal lines (dashed lines). We show SSP models (thick red curves) and CSF models (dotted-dashed blue curves) based on theoretical (T) stellar libraries. Such models only include stellar lines and are not optimized for the stellar \lya. The models correspond to either 5 or 10 Myr, depending on which model fits better. In each panel, we give the KISSR ID, the model age, and the model metallicity. The \lya~emitters are marked with asterisks.}
\label{fig12}
\end{figure}

\begin{figure}
\epsscale{1}
\plotone{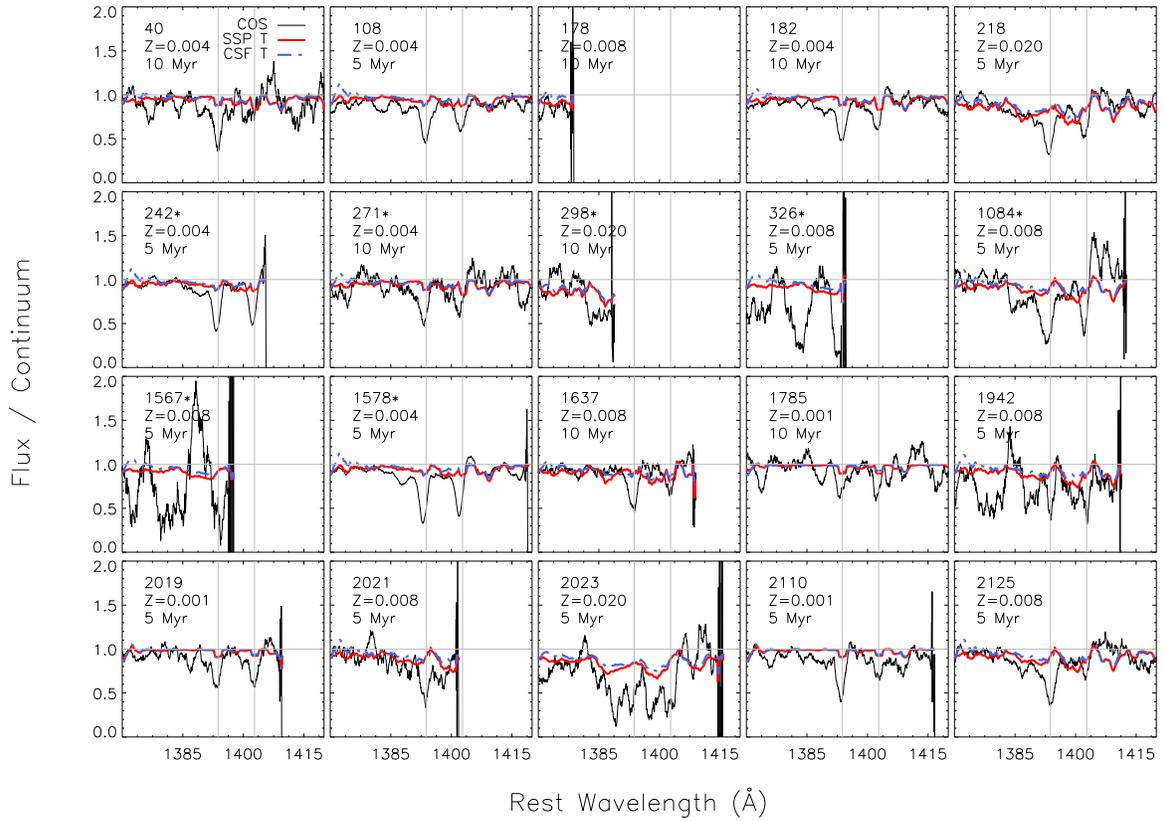}
\caption{Similar to Figure~\ref{fig12} but for the Si\,{\sc iv}~1400 resonance doublets. The vertical gray lines mark the positions of the Si\,{\sc iv} lines.}
\label{fig13}
\end{figure}

\begin{figure}
\epsscale{1}
\plotone{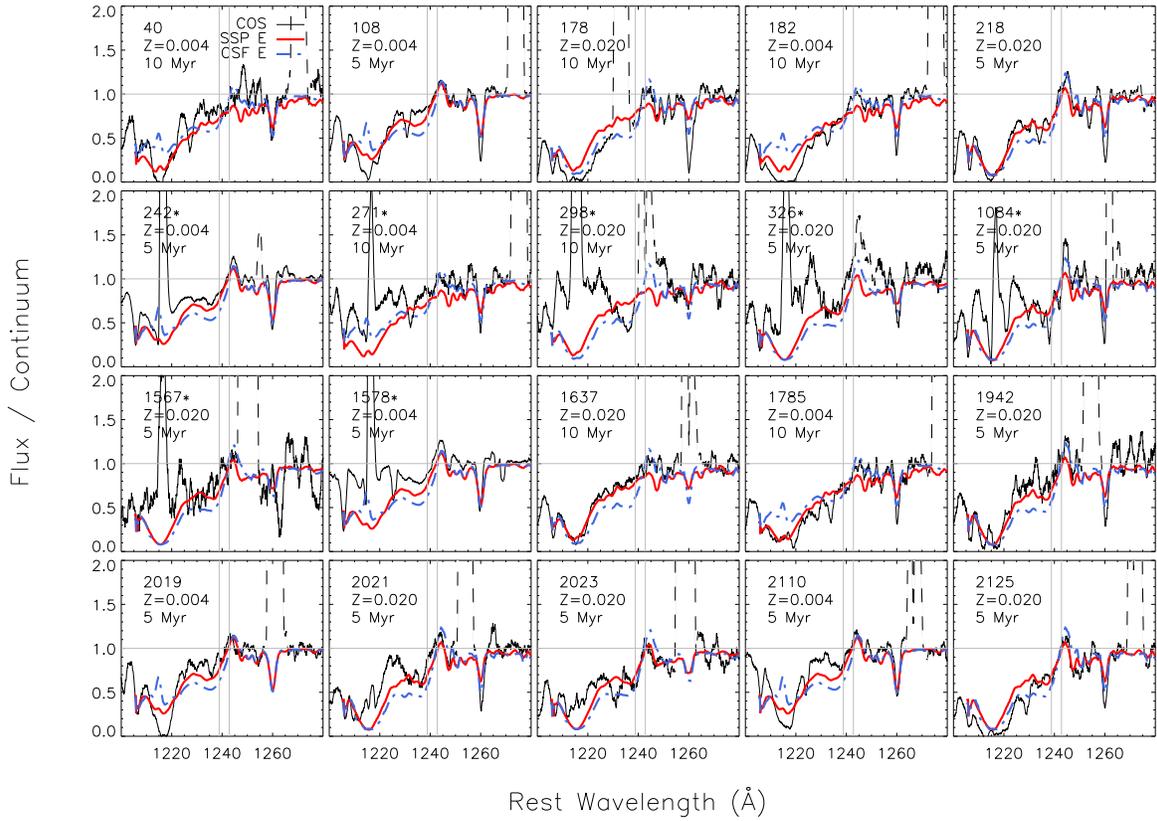}
\caption{Similar to Figure~\ref{fig12} but the models are based on empirical (E) stellar libraries. The empirical models are contaminated with Milky Way lines, in particular, \lya. Only two metallicities are available for the empirical models.}
\label{fig14}
\end{figure}

\begin{figure}
\epsscale{1}
\plotone{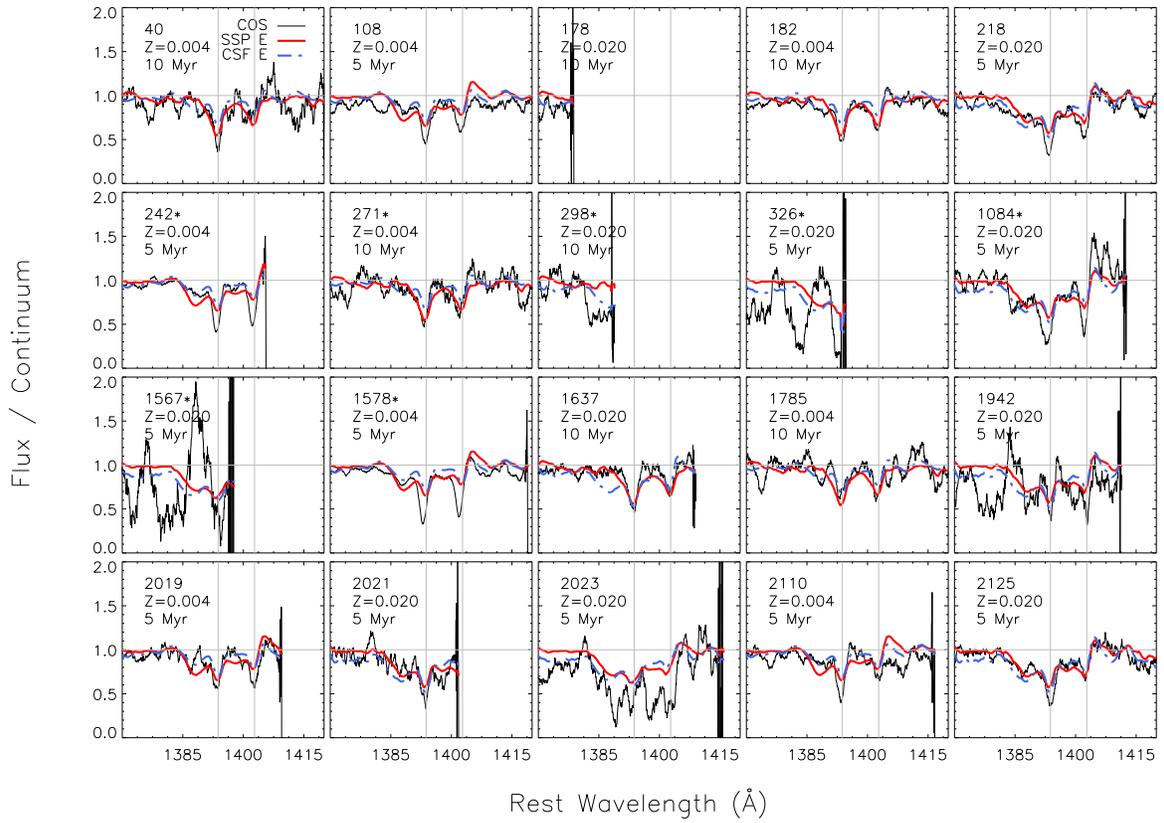}
\caption{Similar to Figure~\ref{fig14} but for the Si\,{\sc iv}~1400 resonance doublets.}
\label{fig15}
\end{figure}

We compared the COS spectra with SSP and CSF models of 5 and 10 Myr. We select this age range because between 5 and 10 Myr, the strength of N\,{\sc v} noticeably decreases and these ages are sufficient for our purpose of demonstrating how young the dominant stellar populations are. We found the best-fit model to the N\,{\sc v} and Si\,{\sc iv} profiles by eye. The comparison was performed on the rectified spectra, as the models are dust-free but the data are not. The metallicity of the model is that most appropriate for the galaxy based on its oxygen abundance\footnote{We adopted stellar evolution tracks with a metallicity $\rm{log}_{10}(Z/Z_\odot)\sim\rm{log}_{10}[(O/H)/(O/H)_\odot]$, where $Z_\odot=0.020$ corresponds to $12+\rm{log}_{10}(O/H)_\odot=8.83$, i.e., to the old solar value of \cite{gre98}.}. Figures~\ref{fig12} and~\ref{fig13} show the comparisons of the observed N\,{\sc v} and Si\,{\sc iv} profiles with the theoretical models. Similarly, Figures~\ref{fig14}  and~\ref{fig15} show the comparisons with the empirical models.

A comparison of the theoretical and empirical models in Figures~\ref{fig12} and~\ref{fig14} shows that at both ages and at all metallicities, N\,{\sc v} is stronger for the CSF models than for the SSP models, although the difference is less at the lowest metallicity (e.g., for 2019 and 2110). Furthermore, the theoretical CSF models yield stronger N\,{\sc v} profiles than the empirical models at similar ages and metallicities (e.g., 40 and 242). Conversely, the theoretical Si\,{\sc iv} profiles show weaker emission components than the empirical ones. These differences do not severely affect our results.

The observed N\,{\sc v}~profiles of \lya~emitters 298, 326, and 1567 are contaminated with geocoronal O\,{\sc i} lines, and the Si\,{\sc iv} profiles of 11 targets, including four of the seven emitters, are either incomplete or have low signal to noise. For this reason, and because the Si\,{\sc iv} profiles from the theoretical models are too weak compared to observations at young ages (e.g., compared to the empirical models), in the rest of our age-dating analysis, we concentrate on the N\,{\sc v} profiles. For targets with clean N\,{\sc v} profiles, we find that the theoretical SSP models tend to be a better fit than the theoretical CSF models, except at the lowest metallicity (e.g., KISSR 2019 and 2110), where the SSP and CSF N\,{\sc v} profiles are very similar. This is true for the \lya~emitters as well as for the \lya~absorbers. One exception is 1084, which is better fitted by a CSF model. In any case, \lya~absorbers such as spiral galaxy KISSR 218 and metal-poor galaxies KISSR 2019 and 2110 have strong N\,{\sc v} profiles and large H$\alpha$ $EW$s indicative of ages younger than $\sim$10 Myr. Thus, they would be expected to have strong \lya~emission.  However, they are absorbers, which is consistent with their low gas-flow velocities (see Table~\ref{tab5}). In addition, one cannot rule out attenuation of \lya~due to the presence of dust. We also find that the emitters have \lya~equivalent widths that are consistent with their young ages, but that are small compared to model predictions for young starbursts. Thus, the emitters also require the presence of scattering in H\,{\sc i} and attenuation by dust. The take away point is that even at young ages ($\lesssim10$\,Myr) targets can show \lya~in absorption if the gas surrounding the starburst is almost static or if the dust column density surrounding the hot stars is significant. 

Table~\ref{tab7} gives the stellar masses and star formation rates derived from the UV spectra by adopting the ages in the second column of the table. For a given age, metallicity, IMF, and star formation history, the theoretical value of $L_{1500}$ only depends of the mass of the stellar population. We give two estimates of the mass and the star formation rate, one corresponding to the SSP and the other to the CSF model. The masses are derived from the comparison of the observed and computed luminosities of the stellar continuum at 1500 \AA~(both dust-free). For the computed luminosity, we use the mean of the theoretical and empirical predictions at the adopted age and metallicity. Note that for the SSP models, it is common practice to define a characteristic star formation rate as the ratio of the stellar mass to the age of the population. This is what we quote in column 5 of Table~\ref{tab7}. For comparison, the table also gives the star formation rate derived from the SDSS H$\alpha$ luminosity (corrected for reddening using the starburst attenuation law of \cite{cal00}) and equation (2) in \cite{ken98}. We recall that the SDSS aperture is very similar to the COS aperture. This is not the case for the KISS data, therefore, we omit the optical SFR for targets without SDSS data. We also recall that the star formation rates are not global values. We find that the SSP models yield higher masses and star formation rates than the CSF models by a factor of  1.4 (median). This is because CSF models predict larger values of $L_{1500}$ than the SSP models at the adopted ages. We also find that the SSP models yield systematically higher star formation rates than those derived using H$\alpha$ by a factor of 5 (median) with a standard deviation of 7. The UV SSP SFRs derived from $L_{1500}$ should be viewed with caution as they are inversely proportional to the ages, which are expected to be underestimated by a factor of a few in cases where N\,{\sc v} is contaminated with geocoronal emission or does not show a strong P-Cygni profile. This is the case for KISSR 298 and 1637, which show the largest disagreement between the optical and the UV SFRs. In addition, the UV SSP SFRs depend on the masses, and Wofford et al. (2011) found systematically higher masses derived from $L_{1500}$ compared to masses derived from optical photometry for individual star clusters. On the other hand, in some cases, the agreement between the optical and UV SFRs is very good. This is the case for KISSR 1578, a target with a prominent N\,{\sc v} doublet and a high S/N spectrum. Unfortunately, only four emitters have H$\alpha$ SFRs, which makes it difficult to study trends between \lya~and the SFR.


\begin{deluxetable}{lcccccc}
\tablecolumns{7}
\tablewidth{0pc}
\tabletypesize{\scriptsize}
\tablecaption{Stellar Population Properties}
\tablehead{Galaxy & Age & Mass & Mass & SFR & SFR & SFR \\
\hfill & \hfill & SSP & CSF & SSP & CSF & H$\alpha$ \\
\hfill & Myr & M$_\odot$ & M$_\odot$ & M$_\odot$ yr$^{-1}$ & M$_\odot$ yr$^{-1}$ & M$_\odot$ yr$^{-1}$ \\
(1) & (2) & (3) & (4) & (5) & (6) & (7) }
\startdata
40	&	10	&	7E+06	&	3E+06		&	0.7	&	0.3	&	0.1	\\
108	&	5	&	7E+06	&	5E+06		&	1.4	&	0.9	&	0.3	\\
178	&	10	&	7E+07	&	3E+07		&	6.8	&	2.7	&	2.6	\\
182	&	10	&	2E+07	&	6E+06		&	1.6	&	0.6	&	0.3	\\
218	&	5	&	4E+07	&	2E+07		&	7.2	&	5.0	&	0.7	\\
242*	&	5	&	9E+07	&	6E+07		&	17.6	&	12.1	&	5.6	\\
271*	&	10	&	1E+07	&	5E+06		&	1.1	&	0.5	&	\nodata	\\
298*	&	10	&	2E+08	&	8E+07		&	19.7	&	7.9	&	0.8	\\
326*	&	5	&	2E+07	&	1E+07		&	3.8	&	2.6	&	\nodata	\\
1084*	&	5	&	9E+07	&	6E+07		&	18.5	&	12.8	&	3.1	\\
1567*	&	5	&	4E+06	&	3E+06		&	0.8	&	0.5	&	\nodata	\\
1578*	&	5	&	3E+07	&	2E+07		&	5.0	&	3.5	&	4.0	\\
1637	&	10	&	6E+07	&	3E+07		&	6.5	&	2.6	&	0.3	\\
1785	&	10	&	5E+06	&	2E+06		&	0.5	&	0.2	&	0.3	\\
1942	&	5	&	1E+07	&	8E+06		&	2.3	&	1.6	&	0.2	\\
2019	&	5	&	5E+06	&	3E+06		&	1.0	&	0.7	&	0.3	\\
2021	&	5	&	3E+07	&	2E+07		&	5.2	&	3.6	&	0.6	\\
2023	&	5	&	7E+06	&	5E+06		&	1.5	&	1.0	&	1.2	\\
2110	&	5	&	3E+06	&	2E+06		&	0.6	&	0.4	&	\nodata	\\
2125	&	5	&	1E+07	&	1E+07		&	2.9	&	2.0	&	1.9	\\
Min	&	-	&	3E+06	&	2E+06		&	0.5	&	0.2	&	0.1	\\
Max	&	-	&	2E+08	&	8E+07		&	19.7	&	12.8	&	5.6	\\
Mean	&	-	&	4E+07	&	2E+07		&	5.2	&	3.1	&	1.4	\\
Sigma	&	-	&	5E+07	&	2E+07		&	6.2	&	3.7	&	1.6	\\
\enddata\\[-15pt]
\tablecomments{(1) KISSR ID identifier. The emitters are marked with asterisks. (2) Adopted age. (3) to (6) Stellar population masses and star formation rates corresponding to the COS aperture and derived by comparing the observed and computed UV spectra. The computed spectra correspond to the adopted ages and to SSP or CSF models. (7) Star formation rate derived from the reddening-corrected H$\alpha$ luminosity. The last four rows give the minimum, maximum, mean, and standard deviation for each column.}
\label{tab7}
\end{deluxetable}

\subsection{Wolf-Rayet Stars}

\begin{figure}
\epsscale{1}
\plotone{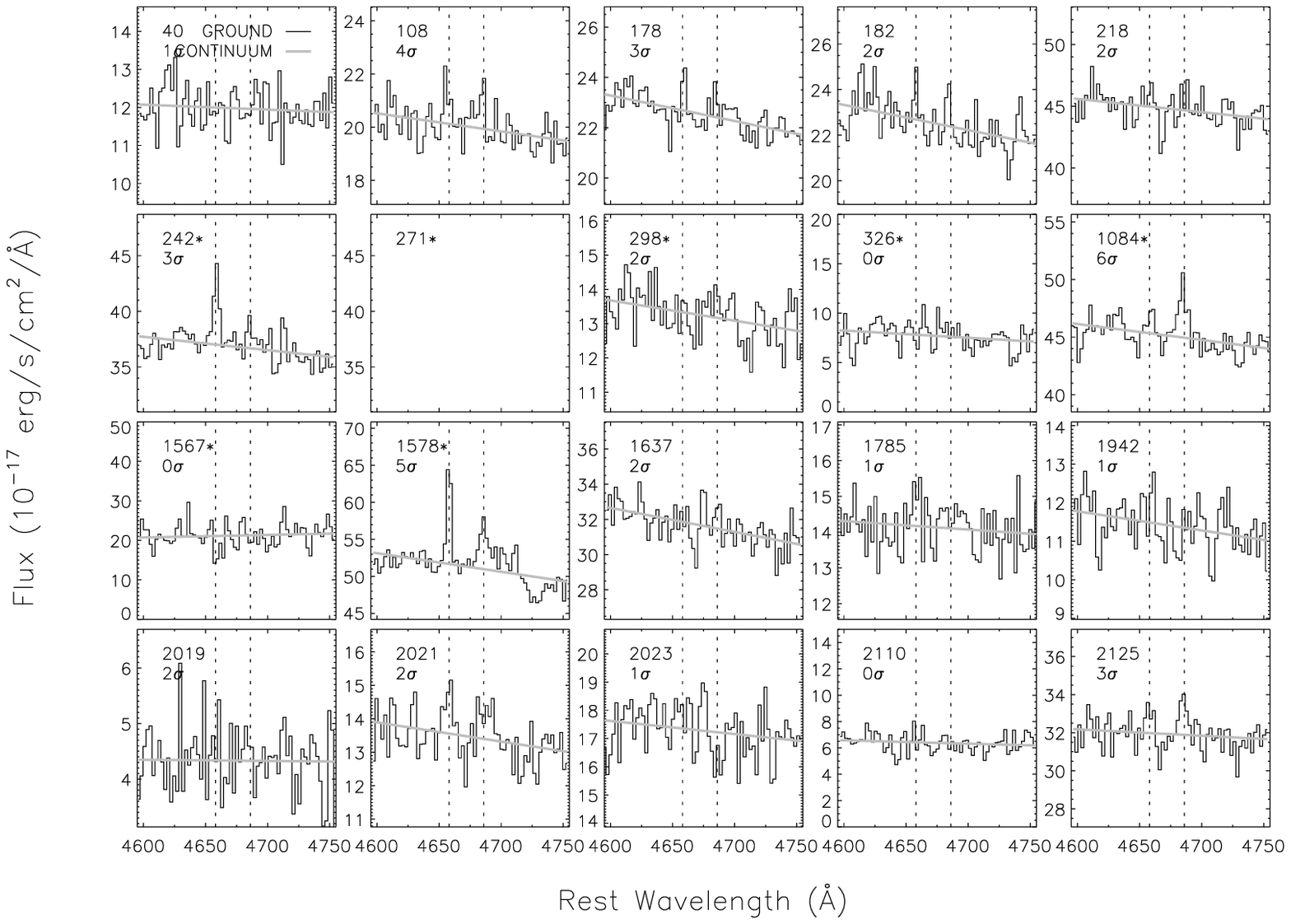}
 \caption{Spectral region around He\,{\sc ii}~$\lambda$4686. The legends give the KISSR ID and the ratios of the amplitude of He\,{\sc ii} to the standard deviation of the nearby continuum. The spectra are from the SDSS DR7 except for galaxies KISSR 326, 1567, and 2110 whose spectra are from Salzer et al. (2005). We do not have an optical spectrum for 271. We degraded the SDSS spectral resolution to match that of the non-SDSS spectra, which is 2.4 \AA. The spectra are corrected for redshift and uncorrected for reddening. The dashed vertical lines mark the positions of the [Fe\,{\sc iii}] $\lambda$4658 and He\,{\sc ii}~$\lambda$4686 emission lines. We overlay in gray a fit to the continuum.}
\label{fig16}
\end{figure}

The detection of broad ($\rm{FWHM}\sim10$ \AA, \citealt{con91}) He\,{\sc ii}~$\lambda$4686 emission (hereafter, He\,{\sc ii}~emission) provides another way of estimating the age of a stellar population. In the absence of an AGN, such emission originates in the dense winds of WR stars \citep{bea29, bib12}, whose progenitors are massive stars (M$_{\rm{initial}}\gtrsim25$, \citealt{mae94}) with lifetimes of $\sim5$ Myr \citep{mey05}.  Other optical WR emission lines are C\,{\sc iii} $\lambda$4650 and N\,{\sc iii} $\lambda$4640, which are generally weaker. For each galaxy, we show in Figure~\ref{fig16} the spectral region around the He\,{\sc ii}~$\lambda$4686 line.  The figure shows that in some cases, the [Fe\,{\sc iii}] $\lambda$4658 peak is taller than the He\,{\sc ii}~$\lambda$4686 peak. As in \cite{lop10}, we used multiple Gaussians to fit the spectral region in the vicinity of 4686 \AA. The Gaussian centers were fixed at wavelengths where lines were expected. We detect He\,{\sc ii} at the $3\sigma$ level or better in six cases, i.e., the amplitude of He\,{\sc ii} is at least three times the standard deviation of the nearby continuum in six cases. In five of the latter cases, the He\,{\sc ii} line is broad, i.e., $\rm{FWHM}\sim10$ \AA. Unfortunately, for four of the seven \lya~emitters, the optical spectrum is either unavailable (KISSR 271), of inadequate spectral resolution (KISSR 326 and KISSR 1567), or very noisy (KISSR 298). The three emitters with adequate optical spectra have He\,{\sc ii} detections with $\rm{FWHM}\ge6$ \AA. This includes KISSR 1084, which may be of composite spectral type, and whose broad He\,{\sc ii} emission is accompanied by narrow/nebular He\,{\sc ii} emission. Note that for the latter galaxy, the He\,{\sc ii} peak is taller than the [Fe\,{\sc iii}] $\lambda$4658 peak, unlike in most galaxies with broad He\,{\sc ii} emission. On the other hand, two of the non-\lya~emitters have He\,{\sc ii} detections and $\rm{FWHM}\ge7$ \AA. The lack of \lya~emission in galaxies with broad He\,{\sc ii}  emission could be due to the lack of significant gas flow in the direction towards the observer. This is the case of KISSR 178 and the highly inclined spiral KISSR 2125.  Finally, the non-emitter KISSR 108 only shows narrow He\,{\sc ii} emission. \cite{shi12} interpret the presence of narrow He\,{\sc ii} emission without a broad component, as WR stars that are offset from the emitting gas.  


\begin{deluxetable}{lccc}
\tablecolumns{4}
\tablewidth{0pc}
\tabletypesize{\scriptsize}
\tablecaption{Helium II 4686 \AA.}
\tablehead{Galaxy & FWHM & EW & Comment\\
\hfill & 4686, \AA & 4686, \AA & \hfill \\
(1) & (2) & (3) & (4)}
\startdata
40 & \nodata & \nodata  & \nodata \\
108   & 2 & 0.38$\pm$0.06  & narrow  \\
178   & 7 & 0.45$\pm$0.13   & \nodata \\
182   & \nodata & \nodata  & \nodata \\
218   & \nodata & \nodata   & \nodata \\
242*   & 6 & 0.54$\pm$0.21  & \nodata  \\
271*   & \nodata & \nodata   & no optical \\
298*  & \nodata & \nodata    & \nodata \\
326*   & \nodata & \nodata  & no SDSS\\
1084*  & 10 & 1.19$\pm$0.04  & broad + narrow \\
1567*   & \nodata & \nodata  & no SDSS\\
1578*   & 9 & 0.84$\pm$0.15  & \nodata \\
1637   & \nodata & \nodata  & \nodata \\
1785  & \nodata & \nodata  & \nodata \\
1942   & \nodata & \nodata  & \nodata \\
2019  & \nodata & \nodata  & \nodata \\
2021  & \nodata & \nodata  & \nodata \\
2023   & \nodata & \nodata & \nodata \\
2110   & \nodata & \nodata  & no SDSS\\
2125   & 10 & 1.26$\pm$0.23  & \nodata \\
Min   & 2 & 0.38$\pm$0.06  & \nodata \\
Max  & 10 & 1.26$\pm$0.23  & \nodata \\
\enddata\\[-15pt]
\tablecomments{(1) KISSR ID. We mark the \lya~emitters with asterisks. (2)  \& (3) FWHM and EW of He\,{\sc ii} $\lambda$4686 for galaxies with an SDSS spectrum and a $3\sigma$ detection. The last column indicates if the target has an optical spectrum or if a narrow He\,{\sc ii} emission component is present. The last two rows give the minimun and maximum values in each column.}
\label{tab8}
\end{deluxetable}

\begin{figure}
\epsscale{1}
\plotone{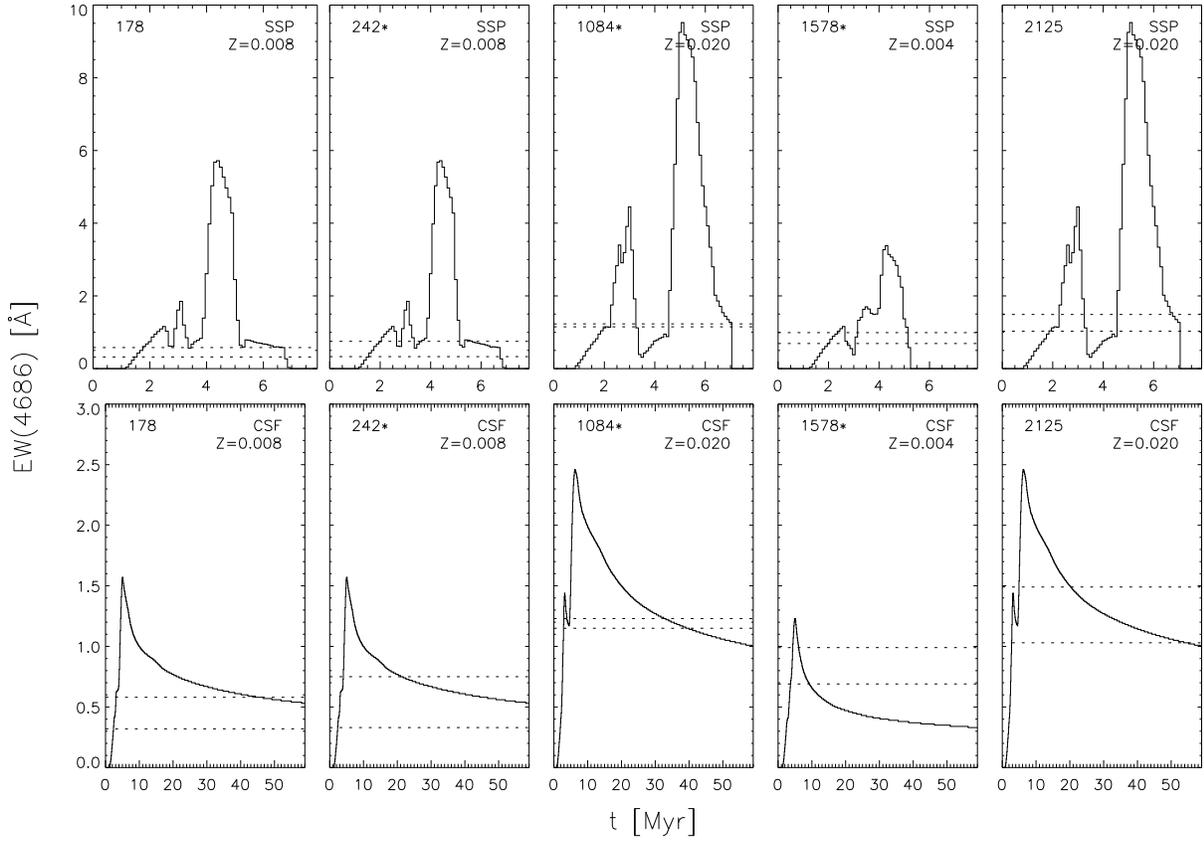}
\caption{He\,{\sc ii}~4686 equivalent width versus time and metallicity for SSP (top) and CSF (bottom) models. In each panel, the dotted horizontal lines represent the upper and lower limits of $EW$(He\,{\sc ii}) for the galaxies with broad He\,{\sc ii}~emission. We give the KISSR ID, SFH, and metallicity in each panel. The emitters are marked with asterisks.}
\label{fig17}
\end{figure}

We compared the observed values of $EW$(He\,{\sc ii}) with Starburst99 \citep{lei99, lei10, vaz05} model predictions. Note that the observed values constitute lower limits due to the dilution of the He\,{\sc ii} line in the stellar continuum of older stellar generations. Table~\ref{tab8} lists FWHM(He\,{\sc ii}) and $EW$(He\,{\sc ii}). Figure~\ref{fig17} shows the comparison of observed and computed values of  $EW$(He\,{\sc ii}) corresponding to single stellar populations (SSP) and continuous star formation (CSF) at the metallicities of the galaxies. Although Figure~\ref{fig17} shows that a given value of $EW$(He\,{\sc ii}) can be reached at multiple times, the range of times is restricted to $10^6-10^7$ yr for the SSP models, and $10^6-10^8$ yr for the CSF models. The presence of broad He\,{\sc ii}~emission rules out an IMF deficient in massive stars or a population dominated by late-B and A stars. Therefore, according to the predictions by \cite{cha93} one would expect positive values of $EW$(\lya) for the galaxies with broad He\,{\sc ii} emission. As previously mentioned, the lack of \lya~emission in KISSR 178 and KISSR 2125 is probably due to the dust geometry and inclination.  In conclusion, we find no one to one correspondence between the \lya~emission and the presence of WR stars. This is in agreement with our age-dating analysis based on the N\,{\sc v} profiles, where we concluded that some of the galaxies with young stellar populations are absorbers.


\section{Discussion}

\subsection{\lya~and Gas Flows}

 \cite{kun98} analyzed \hst/GHRS observations of 8 nearby \hii~galaxies covering metallicities from 12+log(O/H)=8.0 to solar, and reddenings from E(B-V)=0.1 to 0.55. They found velocity offsets between the \hi~and the \hii~gas of up to 200 km s$^{-1}$ in the four galaxies with Ly$\alpha$ emission (P-Cygni profiles), whereas they found broad damped Ly$\alpha$ absorptions at the velocity of the \hii~gas in galaxies with no signs of \hi~outflows from the LIS lines. These authors concluded that outflows are the main determining factor of the Ly$\alpha$ escape from galaxies. Our kinematical results are in very good agreement with these findings. However, one still needs to assess the relative importance of scattering in \hi~and destruction by dust on the \lya~escape. 
 
We also find that the \lya~redshift is higher than the blueshift of the LIS lines by a factor of $1.5-2$. This is in agreement with the high redshift observations of \cite{sha03} and the expanding-shell radiation transport model of \cite{ver06}.  
 
Finally, note that the \lya~troughs are the result of a superposition of \lya~components. Therefore, the dominant component does not need to be at zero velocity. It is interesting that for our targets and for the targets in \cite{kun98}, the \lya~absorption centroid is consistent with the velocity of the \hii~gas. 
 
\subsection{KISSR 298}

KISSR 298 is the double-peaked \lya~emitter.  Such a profile is generally interpreted as a signature of \lya~radiation transfer through a non-static medium (see references given in section 4.5 of \cite{ver08}. For this object, we find an expansion velocity of $150\pm47$ km s$^{-1}$ from the LIS lines. This is high compared to the expansion velocities for the two double-peaked spectra in \cite{ver08}, which are almost consistent with a static medium (they are $<25$ km s$^{-1}$). 

In KISSR 298, the blueshifted peak is twice as strong as the redshifted peak. Stronger blue peaks have been observed in star-forming galaxies at redshifts of $z=2-3$ \citep{ver07,kul12}. They are qualitatively consistent with an inflow \citep{ver06} and could be due to in-falling gas originating in a galactic fountain, as in the model of \cite{ten96}. The fact that in this system the dip between peaks essentially reaches the continuum is an interesting feature. Another possibility is that KISSR 298 has a broad \lya~emission line blended with an absorption, and that the absorption center is shifted redward. The shift could be caused by the relative motion of the emission region and the absorbing gas. A system corresponding to this description is shown in panel f of figure 3 of \cite{wil05}, which corresponds to the second cloud from the top in their figure 4.  Unfortunately, the metal absorption lines of KISSR 298 are noisy and it is hard to tell if there is are unexpected absorptions near the wavelength corresponding to the \lya~"absorption". As can be seen in Figure~\ref{fig11}, the point equidistant to the two \lya~peaks of KISSR 298, which is marked with a dashed line, has a velocity coincident with the Si\,{\sc ii} 1304 and C\,{\sc ii}~$\lambda$1336 absorptions. 

\subsection{\lya~versus Aperture Size and Distance}

We checked if there is an observational bias due to distance among the \lya~emitters in our sample. We excluded KISSR 1084 and 1567 from the analysis as one is of composite type and for the other the TA failed. In order to separate the effect of distance and metallicity/reddening, we divided the emitters in two sets: the low metallicity irregulars (KISSR 242, 1578, and 271) and the higher metallicity spirals (KISSR 298 and 326). Each set spans a factor of about 2 in distance (1.75  for the spirals and 1.7 for the irregulars). The spirals have low inclinations. Therefore the inclination angle is a secondary effect. We find that $L(1500)$, $EW(\rm{Ly}\alpha)$, and $L(Ly\alpha)$ simultaneously increase with distance for both metallicity sets (see Table~\ref{tab6}). Note that this result is based on small number statistics and that the general trend at much higher redshifts is that brighter objects have lower equivalent widths (e.g. \citealt{and06}). 

For the emitters, we find that $EW(Ly\alpha)$ is low compared to values determined in other studies at similar redshifts with larger apertures. \cite{gia96} found that $EW(Ly\alpha)=30$ \AA~on average for a sample of galaxies observed through the $10''\times20''$ aperture of \textit{IUE}. We attribute this to our smaller aperture. Indeed, as previously mentioned, \cite{ost09} found that in nearby galaxies, the bulk of the \lya~emission is in the diffuse component of the galaxies.  We also find that the emitters in our sample have $EW(Ly\alpha)$ slightly below the completeness limit of the $z=0.3$ sample of \cite{cow10}, which is 15 \AA. However, our sample has a median absolute \textit{B} magnitude of -19.4 (see Table~\ref{tab1}), which is similar to that of the $z=0.3$ sample (see their figure 23). This again could be an aperture effect. For Lyman-Break Galaxies (LBGs) at $z\sim3$, the median value of $EW(Ly\alpha)$ is zero (see figure 8 of \citealt{sha03}). Therefore, the $EWs$ of the emitters in our sample are consistent with the latter median. 

\subsection{Ly$\alpha$ versus Metallicity/Reddening}

If a homogeneous \hi~layer covers most of the \hii~region, the value of $EW($\lya) is expected to depend on the dust extinction because the \lya~line flux is more strongly reduced (due to multiple scattering effects) than the adjacent continuum. 

Reddening due to the presence of dust is expected to increase with metallicity. We use oxygen as a gauge of metallicity. We re-derived the oxygen abundances of the 16 galaxies with SDSS spectra using the method described in \cite{sal05}. In summary, in cases where the weak [O\,{\sc iii}] $\lambda$4363 line was not detected, we used the mean of the oxygen abundances derived from the EP and SDSS COARSE calibrations, and in cases where the R23 abundance could be computed, we used the mean of the EP, SDSS and R23 abundances (all methods are defined in \citealt{sal05}). The oxygen abundances from the KISSR and the SDSS data are in agreement within the uncertainties and the final values are those given in Table~\ref{tab1}.

We find that the two most metal-poor and least reddened galaxies, KISSR 2019 and 2110, show \lya~in pure absorption, while the most metal-rich and most reddened galaxy, KISSR 298, has the highest value of $EW(Ly\alpha)$. Strong \lya~emission is expected from KISSR 2019 and 2110 based on the strength of their N\,{\sc v} profiles and their high values of $EW(H\alpha)$ ($\sim500$ \AA), which are characteristic of a young burst of star formation. However, these two targets show \lya~in absorption. This is not surprising given that this is also the case for I Zw 18, the most metal-poor galaxy in the local universe \citep{ate09}. Although we suffer from small number statistics, the lack of one to one correspondence between the strength of \lya~and metallicity/reddening is in agreement with the work of \cite{gia96}, which is based on larger optical and ultraviolet apertures than what we use in the present study. Therefore, it is not clear that larger aperture measurements yield a different result regarding \lya~versus metallicity. Our upcoming \lya~plus H$\alpha$/H$\beta$ ratio maps of three of the galaxies in our sample should help establish the role of dust and metallicity in regulating the \lya~escape. These maps will be obtained as part of \hst~GO-12951 and will enable the measurement of spatially resolved and integrated \lya~and reddening values. 

\section{SUMMARY AND CONCLUSION}

\begin{enumerate}
\item We used the~\hst~COS to observe 20 nearby ($z\sim0.03$) galaxies spanning a broad range in properties (Table~\ref{tab1}). Our data covers the wavelength range $1138-1457$ \AA~and samples circular regions of 0.9-2.4 kpc in radius within each galaxy. We studied correlations between the \lya~line properties and galaxy properties including distance, metallicity, reddening, starburst phase, and gas flow properties. Our study includes the analysis of ancillary optical spectroscopy from SDSS and KISS. Based on the BPT diagram (Figure~\ref{fig1}), the galaxies are dominated by star-formation except for KISSR 1084, which is of composite spectral type.
\item We found \lya~profile shapes representative of what has been observed between redshifts of $z\sim0$ and $z\sim5$ (e.g., \citealt{kun98,tap07,kul12}), including strong absorptions, P-Cygni profiles with redshifted emission, and double emission (Figure~\ref{fig10}). 
\item Seven out of the 20 galaxies have net \lya~emission, including spiral and irregular galaxies and composite galaxy KISSR 1084 (Figures~\ref{fig3} and~\ref{fig10}). For these seven emitters, the \lya~equivalent width is in the range $1-12$~\AA~and the range of escape fractions within the COS aperture is  $1-12$ \% (Table~\ref{tab6}). The comparison with escape fractions at higher redshifts is complicated by the fact that at higher redshifts, observations include more if not all of the galaxy in the aperture.
\item A face-on spiral, KISSR 298, shows two peaks of \lya~emission separated by 370 km s$^{-1}$, one blueshifted and one redshifted and half as strong. Unfortunately, we were unable to determine if this profile is a strong emission with absorption within, as LIS lines for this galaxy are too noisy.
\item Excluding KISSR 298, the emitters have \lya~peaks redshifted to $172\pm49$ km s$^{-1}$ with respect to H$\alpha$ (mean $\pm$ standard deviation). Including all emitters, the mean O\,{\sc i} $\lambda$1302 and C\,{\sc ii} $\lambda$1334 absorption blueshifts are $-117\pm47$ and $-94\pm20$ km s$^{-1}$. Therefore, excluding KISSR 298, the \lya~redshift is larger than the interstellar absorption blueshift by approximately a factor of $1.5-2$, as found at high redshift (e.g. Shapley et al. 2003), and in agreement with radiation transfer models (e.g., Verhamme et al. 2006). 
\item For the absorbers, \lya~is centered at $0\pm38$ km s$^{-1}$ and O\,{\sc i} and  C\,{\sc ii} have centroids at $-23\pm28$ and $-21\pm31$ km s$^{-1}$. Thus, most absorbers are consistent with having static or low velocity \hi~gas. This supports earlier findings (e.g. of Kunth et al. 1998).
\item The outflow velocity of Si {\sc iii} $\lambda$1206 is typically, $30-50$ km s$^{-1}$ greater than that of the \hi~gas (Table~\ref{tab5}).
\item We found no one to one correspondence between the strength of \lya~and the reddening/metallicity measured within the COS aperture (Tables~\ref{tab6} and~\ref{tab7}). In particular, we found three low-inclination spirals each showing different \lya~profiles, i.e., pure absorption, double emission, and P-Cygni profile, and the two most metal-poor and least-reddened galaxies in our sample are pure absorbers.
\item We find stellar populations consistent with SSPs of ages in the range 5-10 Myr among the \lya~emitters and the \lya~absorbers. Similarly, we detect signatures of the presence of Wolf-Rayet stars among the emitters and the absorbers. Therefore, a young starburst does not guarantee strong \lya~emission.
\item In conclusion, a picture emerges where the \lya~photons escape through regions of low \hi~and dust column densities, where gas outflows are occurring.
\item In \hst~cycle 20, we will map three low-inclination spirals from our sample in the \lya~line and the H$\beta$/H$\alpha$ ratio (proxy for dust attenuation). This will make it possible to study the relative importance of gas flows and dust content in regulating the \lya~escape.

\end{enumerate}

\section{Acknowledgments}
Support for this work has been provided by NASA through an award to University of Colorado (Boulder) entitled "Cosmic Origins Spectrograph GTO". Award number NNX08AC14G. We thank C. Danforth, K. France, and J. Tumlinson for their codes, which were used for reducing (CD) and analyzing (KF and JT) the COS data. We also thank C. Thom for his help with the \hi~21 cm data, and Z. Zheng for helpful comments regarding the \lya~profile of KISSR 298. Finally, we thank the referee for comments that have helped to greatly improve this paper.



\end{document}